\documentclass[journal]{IEEEtran}
\IEEEoverridecommandlockouts

\usepackage{fancyhdr}
\usepackage{epsfig}
\usepackage{threeparttable}
\usepackage{epsf,epsfig}
\usepackage{amsthm}
\usepackage[cmex10]{amsmath}
\usepackage{amssymb, amsfonts}
\usepackage[noadjust]{cite}
\usepackage{dsfont}
\usepackage{subfigure}
\usepackage{color}
\usepackage{soul}
\usepackage{amssymb}
 \usepackage{accents}
  \usepackage{algorithm}
   \usepackage[english]{babel}
    \usepackage{url}
    \usepackage{framed} 
        \usepackage{bm}

\begin{document}
\newtheorem{theorem}{Theorem}
\newtheorem{acknowledgement}[theorem]{Acknowledgement}
\newtheorem{axiom}[theorem]{Axiom}
\newtheorem{case}[theorem]{Case}
\newtheorem{claim}[theorem]{Claim}
\newtheorem{conclusion}[theorem]{Conclusion}
\newtheorem{condition}[theorem]{Condition}
\newtheorem{conjecture}[theorem]{Conjecture}
\newtheorem{criterion}[theorem]{Criterion}
\newtheorem{definition}{Definition}
\newtheorem{exercise}[theorem]{Exercise}
\newtheorem{lemma}{Lemma}
\newtheorem{corollary}{Corollary}
\newtheorem{notation}[theorem]{Notation}
\newtheorem{problem}[theorem]{Problem}
\newtheorem{proposition}{Proposition}
\newtheorem{scheme}{Scheme}   
\newtheorem{solution}[theorem]{Solution}
\newtheorem{summary}[theorem]{Summary}
\newtheorem{assumption}{Assumption}
\newtheorem{example}{\bf Example}
\newtheorem{remark}{\bf Remark}

\def\qed{$\Box$}
\def\QED{\mbox{\phantom{m}}\nolinebreak\hfill$\,\Box$}
\def\proof{\noindent{\emph{Proof:} }}
\def\poof{\noindent{\emph{Sketch of Proof:} }}
\def
\endproof{\hspace*{\fill}~\qed
\par
\endtrivlist\unskip}
\def\endproof{\hspace*{\fill}~\qed\par\endtrivlist\vskip3pt}

\def\E{\mathsf{E}}
\def\eps{\varepsilon}
\def\phi{\varphi}
\def\Lsp{{\boldsymbol L}}
\def\Bsp{{\boldsymbol B}}
\def\lsp{{\boldsymbol\ell}}
\def\Ltsp{{\Lsp^2}}
\def\Lpsp{{\Lsp^p}}
\def\Linsp{{\Lsp^{\infty}}}
\def\LtR{{\Lsp^2(\Rst)}}
\def\ltZ{{\lsp^2(\Zst)}}
\def\ltsp{{\lsp^2}}
\def\ltZt{{\lsp^2(\Zst^{2})}}
\def\ninN{{n{\in}\Nst}}
\def\oh{{\frac{1}{2}}}
\def\grass{{\cal G}}
\def\ord{{\cal O}}
\def\dist{{d_G}}
\def\conj#1{{\overline#1}}
\def\ntoinf{{n \rightarrow \infty }}
\def\toinf{{\rightarrow \infty }}
\def\tozero{{\rightarrow 0 }}
\def\trace{{\operatorname{trace}}}
\def\ord{{\cal O}}
\def\UU{{\cal U}}
\def\rank{{\operatorname{rank}}}
\def\acos{{\operatorname{acos}}}

\def\SINR{\mathsf{SINR}}
\def\SNR{\mathsf{SNR}}
\def\SIR{\mathsf{SIR}}
\def\tSIR{\widetilde{\mathsf{SIR}}}
\def\Ei{\mathsf{Ei}}
\def\l{\left}
\def\r{\right}
\def\({\left(}
\def\){\right)}
\def\lb{\left\{}
\def\rb{\right\}}

\setcounter{page}{1}

\newcommand{\eref}[1]{(\ref{#1})}
\newcommand{\fig}[1]{Fig.\ \ref{#1}}

\def\bydef{:=}
\def\ba{{\mathbf{a}}}
\def\bb{{\mathbf{b}}}
\def\bc{{\mathbf{c}}}
\def\bd{{\mathbf{d}}}
\def\bee{{\mathbf{e}}}
\def\bff{{\mathbf{f}}}
\def\bg{{\mathbf{g}}}
\def\bh{{\mathbf{h}}}
\def\bi{{\mathbf{i}}}
\def\bj{{\mathbf{j}}}
\def\bk{{\mathbf{k}}}
\def\bl{{\mathbf{l}}}
\def\bn{{\mathbf{n}}}
\def\bo{{\mathbf{o}}}
\def\bp{{\mathbf{p}}}
\def\bq{{\mathbf{q}}}
\def\br{{\mathbf{r}}}
\def\bs{{\mathbf{s}}}
\def\bt{{\mathbf{t}}}
\def\bu{{\mathbf{u}}}
\def\bv{{\mathbf{v}}}
\def\bw{{\mathbf{w}}}
\def\bx{{\mathbf{x}}}
\def\by{{\mathbf{y}}}
\def\bz{{\mathbf{z}}}
\def\b0{{\mathbf{0}}}

\def\bA{{\mathbf{A}}}
\def\bB{{\mathbf{B}}}
\def\bC{{\mathbf{C}}}
\def\bD{{\mathbf{D}}}
\def\bE{{\mathbf{E}}}
\def\bF{{\mathbf{F}}}
\def\bG{{\mathbf{G}}}
\def\bH{{\mathbf{H}}}
\def\bI{{\mathbf{I}}}
\def\bJ{{\mathbf{J}}}
\def\bK{{\mathbf{K}}}
\def\bL{{\mathbf{L}}}
\def\bM{{\mathbf{M}}}
\def\bN{{\mathbf{N}}}
\def\bO{{\mathbf{O}}}
\def\bP{{\mathbf{P}}}
\def\bQ{{\mathbf{Q}}}
\def\bR{{\mathbf{R}}}
\def\bS{{\mathbf{S}}}
\def\bT{{\mathbf{T}}}
\def\bU{{\mathbf{U}}}
\def\bV{{\mathbf{V}}}
\def\bW{{\mathbf{W}}}
\def\bX{{\mathbf{X}}}
\def\bY{{\mathbf{Y}}}
\def\bZ{{\mathbf{Z}}}

\def\mA{{\mathbb{A}}}
\def\mB{{\mathbb{B}}}
\def\mC{{\mathbb{C}}}
\def\mD{{\mathbb{D}}}
\def\mE{{\mathbb{E}}}
\def\mF{{\mathbb{F}}}
\def\mG{{\mathbb{G}}}
\def\mH{{\mathbb{H}}}
\def\mI{{\mathbb{I}}}
\def\mJ{{\mathbb{J}}}
\def\mK{{\mathbb{K}}}
\def\mL{{\mathbb{L}}}
\def\mM{{\mathbb{M}}}
\def\mN{{\mathbb{N}}}
\def\mO{{\mathbb{O}}}
\def\mP{{\mathbb{P}}}
\def\mQ{{\mathbb{Q}}}
\def\mR{{\mathbb{R}}}
\def\mS{{\mathbb{S}}}
\def\mT{{\mathbb{T}}}
\def\mU{{\mathbb{U}}}
\def\mV{{\mathbb{V}}}
\def\mW{{\mathbb{W}}}
\def\mX{{\mathbb{X}}}
\def\mY{{\mathbb{Y}}}
\def\mZ{{\mathbb{Z}}}

\def\cA{\mathcal{A}}
\def\cB{\mathcal{B}}
\def\cC{\mathcal{C}}
\def\cD{\mathcal{D}}
\def\cE{\mathcal{E}}
\def\cF{\mathcal{F}}
\def\cG{\mathcal{G}}
\def\cH{\mathcal{H}}
\def\cI{\mathcal{I}}
\def\cJ{\mathcal{J}}
\def\cK{\mathcal{K}}
\def\cL{\mathcal{L}}
\def\cM{\mathcal{M}}
\def\cN{\mathcal{N}}
\def\cO{\mathcal{O}}
\def\cP{\mathcal{P}}
\def\cQ{\mathcal{Q}}
\def\cR{\mathcal{R}}
\def\cS{\mathcal{S}}
\def\cT{\mathcal{T}}
\def\cU{\mathcal{U}}
\def\cV{\mathcal{V}}
\def\cW{\mathcal{W}}
\def\cX{\mathcal{X}}
\def\cY{\mathcal{Y}}
\def\cZ{\mathcal{Z}}
\def\cd{\mathcal{d}}
\def\Mt{M_{t}}
\def\Mr{M_{r}}
\def\O{\Omega_{M_{t}}}
\newcommand{\figref}[1]{{Fig.}~\ref{#1}}
\newcommand{\tabref}[1]{{Table}~\ref{#1}}

\newcommand{\var}{\mathsf{var}}
\newcommand{\fb}{\tx{fb}}
\newcommand{\nf}{\tx{nf}}
\newcommand{\BC}{\tx{(bc)}}
\newcommand{\MAC}{\tx{(mac)}}
\newcommand{\Pout}{p_{\mathsf{out}}}
\newcommand{\nnn}{\nn\\}
\newcommand{\FB}{\tx{FB}}
\newcommand{\TX}{\tx{TX}}
\newcommand{\RX}{\tx{RX}}
\renewcommand{\mod}{\tx{mod}}
\newcommand{\m}[1]{\mathbf{#1}}
\newcommand{\td}[1]{\tilde{#1}}
\newcommand{\sbf}[1]{\scriptsize{\textbf{#1}}}
\newcommand{\stxt}[1]{\scriptsize{\textrm{#1}}}
\newcommand{\suml}[2]{\sum\limits_{#1}^{#2}}
\newcommand{\sumlk}{\sum\limits_{k=0}^{K-1}}
\newcommand{\eqhsp}{\hspace{10 pt}}
\newcommand{\tx}[1]{\texttt{#1}}
\newcommand{\Hz}{\ \tx{Hz}}
\newcommand{\sinc}{\tx{sinc}}
\newcommand{\tr}{\mathrm{tr}}
\newcommand{\diag}{\mathrm{diag}}
\newcommand{\MAI}{\tx{MAI}}
\newcommand{\ISI}{\tx{ISI}}
\newcommand{\IBI}{\tx{IBI}}
\newcommand{\CN}{\tx{CN}}
\newcommand{\CP}{\tx{CP}}
\newcommand{\ZP}{\tx{ZP}}
\newcommand{\ZF}{\tx{ZF}}
\newcommand{\SP}{\tx{SP}}
\newcommand{\MMSE}{\tx{MMSE}}
\newcommand{\MINF}{\tx{MINF}}
\newcommand{\RC}{\tx{MP}}
\newcommand{\MBER}{\tx{MBER}}
\newcommand{\MSNR}{\tx{MSNR}}
\newcommand{\MCAP}{\tx{MCAP}}
\newcommand{\vol}{\tx{vol}}
\newcommand{\ah}{\hat{g}}
\newcommand{\tg}{\tilde{g}}
\newcommand{\teta}{\tilde{\eta}}
\newcommand{\heta}{\hat{\eta}}
\newcommand{\uh}{\m{\hat{s}}}
\newcommand{\eh}{\m{\hat{\eta}}}
\newcommand{\hv}{\m{h}}
\newcommand{\hh}{\m{\hat{h}}}
\newcommand{\Po}{P_{\mathrm{out}}}
\newcommand{\Poh}{\hat{P}_{\mathrm{out}}}
\newcommand{\Ph}{\hat{\gamma}}
\newcommand{\mat}[1]{\begin{matrix}#1\end{matrix}}
\newcommand{\ud}{^{\dagger}}
\newcommand{\C}{\mathcal{C}}
\newcommand{\nn}{\nonumber}
\newcommand{\nInf}{U\rightarrow \infty}

\title{\huge Privacy For Free: Wireless Federated Learning Via\\
Uncoded Transmission With Adaptive Power Control}

\author{Dongzhu Liu and Osvaldo Simeone
\thanks{\noindent The authors are with King's Communications, Learning, and Information Processing (KCLIP) lab at the Department of Engineering of Kings College London, UK (emails: dongzhu.liu@kcl.ac.uk, osvaldo.simeone@kcl.ac.uk). The authors have received funding from the European Research Council (ERC) under the European Unions Horizon 2020 Research and Innovation Programme (Grant Agreement No. 725731).}}

\maketitle

\begin{abstract}
Federated Learning (FL) refers to distributed protocols that avoid direct raw data exchange among the participating devices while training for a common learning task. This way, FL can potentially reduce the information on the local data sets that is leaked via communications. In order to provide formal privacy guarantees, however, it is generally necessary to put in place additional masking mechanisms. When FL is implemented in wireless systems via uncoded transmission, the channel noise can directly act as a privacy-inducing mechanism. This paper demonstrates that, as long as the privacy constraint level, measured via differential privacy (DP), is below a threshold
that decreases with the signal-to-noise ratio (SNR), uncoded transmission achieves privacy ``for free", i.e., without affecting the learning performance. More generally, this work studies adaptive power allocation (PA) for distributed gradient descent in wireless FL with the aim of minimizing the learning optimality gap under privacy and power constraints. Both orthogonal multiple access (OMA) and non-orthogonal multiple access (NOMA) transmission with ``over-the-air-computing" are studied, and solutions are obtained in closed form for an offline optimization setting. Furthermore, heuristic online methods are proposed that leverage iterative one-step-ahead optimization. The importance of dynamic PA and the potential benefits of NOMA versus OMA are demonstrated through extensive simulations.
\end{abstract}


\section{Introduction} 
In modern wireless systems, mobile devices generate and store data that can be utilized to train machine learning models \cite{park2019wireless, zhou2019edge, zhu2020toward}.  While data at one device may be insufficient to obtain effective trained solutions, networked devices can benefit from data stored at other devices via communications.
Federated learning (FL) refers to decentralized training protocols that avoid direct data sharing among devices, while exchanging information about the local models  \cite{li2020federated, kairouz2019advances}.  This has the potential benefits of reducing the communication load and of leaking less information about the local data sets at the devices \cite{melis2019exploiting,fredrikson2015model,bhowmick2018protection}.  A well-established measure the privacy of local data sets with respect to disclosed aggregate statistics is differential privacy (DP) \cite{dwork2014algorithmic}.  Typical DP mechanisms randomize the disclosed statistics by adding random noise~\cite{dwork2014algorithmic}.  This creates a trade-off between accuracy and privacy, as determined by the amount of added noise.  

 {\color{black} This paper investigates the idea of letting the channel noise serve as privacy mechanism.  To this end, we focus on uncoded transmission of the gradients using either orthogonal or non-orthogonal protocols, and we analytically demonstrate that for these transmission schemes, privacy may be obtained  ``for free".  This is in the sense that enforcing a DP constraint causes no performance loss with respect to a non-private design as long as the signal-to-noise ratio (SNR) is sufficiently low.  More generally, we introduce a novel optimal closed-form adaptive transmit power control strategy that optimizes the learning performance while ensuring DP requirements.  }

\subsection{Wireless Federated Learning}
A illustrated in Fig.~\ref{Fig: FL system},
typical wireless FL protocols iterate between local adaptation and centralized combining.  Adaptation involves local optimization steps based on a device's data, while central aggregation amounts to averaging operations on the devices' updates. To reduce the time to convergence, recent work has proposed to leverage computations over multiple access channels \cite{nazer2007computation} as a primitive for the global combining step \cite{zhu2019broadband, sery2020analog, amiri2020machine, zhang2020gradient, yang2020federated}.  Accordingly, all devices simultaneously transmit their updates to edge server using uncoded transmission, which are aggregated ``over-the-air" by exploiting the waveform-superposition property of a multi-access channel.  

To further improve the bandwidth efficiency of this non-orthogonal multiple access (NOMA) scheme, devices can pre-process the analog updates via sparsification based dimensionality reduction~\cite{amiri2020machine}. And the learning performance of this approach can be further enhanced by gradient aware power control~\cite{zhang2020gradient} and joint device selection and beamforming design \cite{yang2020federated}.

As a more conventional solution, FL can be implemented using digital coded transmission. Under digital coded transmission and orthogonal multiple access (OMA), quantization of the local gradients has been proposed to trade off communication bandwidth and convergence rate \cite{alistarh2017qsgd}, with each component of the gradient being represented even by a single bit  \cite{bernstein2018signsgd}. More complex quantization schemes include hierarchical quantization via a low-dimensional codebook on a Grassmann manifold \cite{du2020high}. 
With NOMA, reference \cite{chang2020communication} proposes a strategy whereby each device quantizes the gradient based on its informativeness and on the channel condition.  An alternative is to use one-bit quantization followed by BPSK/QPSK modulation at the devices with NOMA, and to estimate the aggregated gradient  at the edger sever using majority voting~\cite{zhu2020one}. 
%

\subsection{Differential Privacy for Federated Learning}

According to its original motivation, FL may have desirable privacy properties since training is conducted in a distributed manner without sharing the raw data.  Nevertheless, the model updates shared by the devices may reveal information about local data.  For example, a malicious server could potentially infer the presence of  an individual data sample from a learnt model by membership inference attack \cite{shokri2017membership} or model inversion attack \cite{fredrikson2015model}.  DP quantifies information leaked about individual data points by measuring the sensitivity of the disclosed statistics to changes in the input data set at a single data point.  DP can be guaranteed by introducing a level of uncertainty into the released model that is sufficient to mask the contribution of any individual data point \cite{dwork2014algorithmic}.   The most typical approach is to add random perturbations, e.g., Gaussian \cite{abadi2016deep}, Laplacian \cite{wu2019value}, or Binomial noise \cite{agarwal2018cpsgd}, to the released statistics.

DP mechanisms have been investigated for FL under the assumption that the edge server is ``honest-but-curious" and that communication is noiseless and unconstrained.  
In \cite{wei2020federated}, Gaussian noise is added to the local model updates, and the power of the Gaussian noise is adapted to ensure a target privacy level. Analysis indicates that there is a tradeoff between convergence rate and privacy protection levels.  
Furthermore, a higher privacy guarantee is achievable if the DP algorithm uses random mini-batches --- the so-called ``privacy amplification
by subsampling" principle \cite{balle2018privacy}.  
Another DP mechanism based on random quantization is explored in \cite{gandikota2019vqsgd, agarwal2018cpsgd}. 

While the work reviewed so far assumes ideal communication, several recent works have appeared that share the common theme of exploiting the channel noise for differentially private FL. In \cite{seif2020wireless}, each device adds Gaussian noise before transmission via NOMA with static power allocation. The superposition property of NOMA is shown not only to provide benefits in terms of efficient gradient aggregation, but also to offer better privacy guarantees. Instead of injecting noise before transmission, an energy efficient approach is to scale down the transmit power \cite{koda2020differentially}.  A digital counterpart of these ideas is proposed in \cite{sonee2020efficient} which uses quantized gradient descent with privacy-inducing binomial noise. The quantization bits and noise parameters are optimized to maximize the convergence rate under channel capacity and privacy constraints.  

{\color{black}All the discussed works assume a simple static power allocation,  not accounting for the fact that channel noise has a different impact on convergence and privacy level. As our analysis demonstrates, channel noise added in the first iterations tends to impact convergence less significantly than the noise added in later iterations, whereas the privacy level depends on a weighted sum of the inverse noise power across the iteration. These properties, captured by compact analytical expressions derived in this paper, are leveraged to define optimization problems that are solved in closed form, yielding significant performance gains over standard static power allocation.}


\subsection{Contributions and Organization}

In this paper, we study differentially private wireless distributed gradient descent via the direct, uncoded, transmission of gradients from devices to edge server. The channel noise is utilized as a privacy preserving mechanism and dynamic power control is separately optimized for OMA and NOMA protocols with the goal of minimizing the learning optimality gap under privacy and power constraints across a given number of {\color{black}communication blocks.}  The main findings and contributions of the paper can be summarized as follows.

\noindent $\bullet$ {\bf Offline optimized power allocation for OMA and NOMA:}
Considering OMA and NOMA separately, we first analyze the convergence rate and privacy requirements for a given number of iterations under uncoded transmission.  The resulting offline optimization problems are shown to be convex programs, and the optimal dynamic power allocation (PA) is obtained in closed form.  The optimal PA is shown to be adaptive across the iterations, outperforming static PA assumed in prior works. The analytical results prove that privacy can be obtained ``for free" as long as the privacy constraint level is below a threshold that decreases with the signal-to-noise (SNR).  We also demonstrate that it is generally suboptimal to devote part of the transmitted power to actively add noise to the local updates. This is unlike the standard scenario with ideal communication, in which adding noise is essential to ensure DP constraints.

\noindent $\bullet$ {\bf Online power allocation scheme:} A heuristic online approach is then proposed that leverages iterative one-step-ahead optimization based on the offline closed-form solutions, predicted channel state information (CSI).  

\noindent $\bullet$ {\bf Experiments:} We provide extensive numerical results that demonstrate the advantages of 
NOMA over conventional OMA protocols under DP constraints.  We note here that these benefits is not a prior evident, since, with NOMA, devices transmit more frequently, and hence may leak more information if power is not properly allocated.

The remainder of the paper is organized as follows. 
Section~\ref{sec: Preliminaries} introduces the models and definitions. 
Section~\ref{sec: OMA} presents the power allocation design for OMA. The design for NOMA is presented in Section~\ref{sec: NOMA}.  Section~\ref{sec:simulation} provides numerical results, followed by conclusions in Section~\ref{sec: conclusions}.

\section{Models and Definitions}\label{sec: Preliminaries}

\begin{figure}[t]
\centering
\includegraphics[width=9cm]{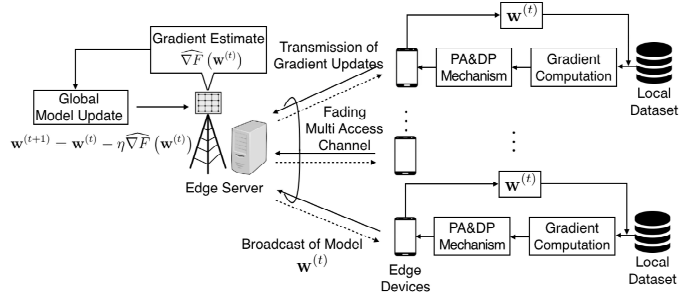}
\vspace{-15pt}
\caption{Differentially private federated edge learning system based on distributed gradient descent.}
\vspace{-15pt}
\label{Fig: FL system}
\end{figure}

As shown in Fig.~\ref{Fig: FL system}, we consider a wireless federated edge learning system comprising a single-antenna edge server and $K$ edge devices connected through it via a shared noisy channel. Each device $k$, equipped with a single antenna, has its own local dataset $\mathcal{D}_k$.  This consists of labelled data samples $\{(\bu_i,v_i)\} \in \mathcal{D}_k$, where $\bu_i$ denotes the vector of covariates and $v_i$ its associated label, which may be continuous or discrete.  Local data sets are disjoint. A common regression or classification model, parameterized by vector $\bw$, is collaboratively trained by the edge devices through communications via edge server.  In this section, we first introduce the learning protocol and the communication model, and then detail which the definition of differential privacy adopted in this work, along with main assumptions.

\subsection{Learning Protocol}
The regularized local loss function for the $k$-th device evaluated at model vector $\bw\in\mathbb{R}^d$ is given by
\begin{align}\label{eq: local loss}
{(\text {Local loss function})} \quad F_k(\bw)=\frac{1}{{D}_k}&\sum_{(\bu,v)\in\mathcal{D}_k}f(\bw;\bu,v) \nn\\
&\quad+\lambda R(\bw),
\end{align}
where $f(\bw;\bu,v)$ is the sample-wise loss function quantifying the prediction error of the model $\bw$ on the training sample $\bu$ with respect to (w.r.t.) its ground-truth label $v$; $D_k=|\mathcal{D}_k|$ is the cardinality of  data set $\mathcal{D}_k$; and $R(\bw)$ is a strongly convex regularization function, which is scaled by hyperparameter $\lambda\geq0$.  The global loss function evaluated at model vector $\bw$ is
\begin{equation}\label{eq: global loss}
{(\text {Global loss function})} \quad F(\bw)=\frac{1}{D_{\sf tot}}\sum_{k=1}^KD_k F_k(\bw),
\end{equation}
where $D_{\sf tot}=\sum_k D_k$.  {\color{black} This amounts to the regularized empirical average of the sample-wise loss functions on the global data set $\mathcal{D}=\bigcup_{k=1}^K \mathcal{D}_k$ obtained as the union of the local data sets.  We note that the training loss (2) is an unbiased estimate of the generalization loss only if the devices observe {independent and identically distributed} (i.i.d.) samples from a common distribution. Nevertheless, this objective is also routinely considered for non-i.i.d. data sets in federated learning \cite{zhao2018federated,wang2019adaptive,reisizadeh2020fedpaq}. Other criteria that may be better suited to account for heterogeneous statistics across the devices may also be considered \cite{li2018federated}, but we leave this aspect for future work.}
The learning process aims to minimize the regularized global loss function as
\begin{equation}\label{eq: LP}
\bw^*=\arg\min F(\bw).
\end{equation}

In order to address problem \eqref{eq: LP}, we study a differentially private implementation of federated distributed gradient descent via gradient-averaging.  As we will detail, privacy is defined here from the point of view of any device with respect to the edge server, which is assumed to be ``honest-but-curious".  Accordingly, the edge server follows the protocol described below, but may attempt to infer information about data at the edge devices.  We do not directly enforce  privacy constraints on the other devices, which are implicitly trusted. More discussion on this point can be found in Section~\ref{sec: conclusions}.

As illustrated in Fig.~\ref{Fig: FL system}, at each $t$-th communication round, {\color{black}with $t=1,\dots, T$, }the edge server broadcasts the current model iterate $\bw^{(t)}$ to all edge devices via the downlink channel.  We assume that downlink communication is ideal, so that each device receives the current model $\bw^{(t)}$ without distortion.  This assumption is practically well justified when the edge sever communicates through a base station with less stringent power constraint than the devices.  
By using the received current model $\bw^{(t)}$ and the local dataset $\mathcal{D}_k$, each device computes the gradient of the local loss function in \eqref{eq: local loss}, that is
\begin{align}\label{eq: local gradient}
{(\text {Local gradient})} \  \nabla F_k\big(\bw^{(t)}\big)=\frac{1}{D_k}&\sum_{(\bu,v)\in\mathcal{D}_k}\nabla f\big(\bw^{(t)};\bu,v\big) \nn\\
&\quad+\lambda \nabla R(\bw^{(t)}).
\end{align}
The devices transmit information about the local gradient \eqref{eq: local gradient} over the wireless shared channel to the edge server.  Based on the received signal, the edge server obtains an estimate $\widehat{\nabla F}\l(\bw^{(t)}\r)$ of the global gradient 
\begin{equation}\label{eq: global gradient}
{(\text {Global gradient})} \quad {\nabla F}\l(\bw^{(t)}\r)=\frac{1}{D_{\sf tot}}\sum_{k=1}^K D_k \nabla F_k\l(\bw^{(t)}\r).
\end{equation}
The edge server then updates the current global model via gradient descent \begin{equation}\label{eq: Model update}
{(\text {Model updating})} \quad \bw^{(t+1)}=\bw^{(t)}-\eta\widehat{\nabla F}\l(\bw^{(t)}\r),
\end{equation}
where $\eta$ denotes the learning rate. The steps in \eqref{eq: local gradient}, \eqref{eq: global gradient}, and \eqref{eq: Model update} are iterated until a convergence condition is met.

The transmission of the gradient \eqref{eq: local gradient}  from each $k$-th device may reveal information about the local data sets to the edge server.  This motivates the use of DP as a rigorous mathematical framework to provide  privacy guarantees that are agnostic to the computing resources and data processing requirements of the edge server.  This will be detailed in Sec.~\ref{sec: def DP}.

\subsection{Communication Model}
All devices communicate via the uplink to the edge server on the shared 
wireless channel using uncoded transmission. {\color{black}The main focus of this paper is the study of uncoded non-orthogonal multiple access (NOMA)\footnote{\color{black}In this context, NOMA is used as a transmission strategy, and it does not imply the use of specific decoders, such as successive decoding.} protocol, which enables over-the-air computing. For reference, we also study orthogonal multiple access (OMA) scheme under the same assumption of uncoded transmission.  We note that it would be useful to include  digital coded strategies for OMA as a benchmark.  However, the design of digital communication protocols under DP constraints is a non-trivial problem that is currently subject to research \cite{sonee2020efficient}. }

 We assume a block flat-fading channel, where the channel coefficients remain constant within a communication block, and they vary in a potentially correlated way over successive blocks. {\color{black}
 Each block contains $d$ channel uses, allowing the uncoded transmission of a gradient vector. Due to memory and processing complexity constraints, on-device machine learning models are typically of small size, so that the model parameters dimension $d$ can be assumed to be limited to a few tens of  thousands of entries  \cite{ravi2019efficient}.  In this case, considering that typical coherence blocks may be of the same order of magnitude \cite{debaenst2020rms,wang2018doppler}, it is generally feasible to communicate the entire gradient vectors within one communication block. For larger model sizes, the gradient would need to be communicated across multiple coherence blocks -- a setting that we leave for future investigations.}
%
{\color{black}We consider total $I$ blocks available for training.}

 As in most papers on over-the-air computing \cite{zhu2019broadband,yang2020federated, amiri2020machine, sery2020analog,seif2020wireless}, we assume perfect channel state information (CSI) at all nodes, so that each device can compensate for the phase of its own channel, ensuring the effective channel $h_k^{[i]}$ for each device $k$ and block $i$ is \emph{real and non-negative}. { \color{black}
 This allows us to focus on a real channel model with non-negative channel gain, and it simplifies the design in power control parameters.  We note that this assumption is also made in \cite{zhang2020gradient,cao2020optimized}, and that, as in these prior works, we do not make optimality claims in this regard.} Details for OMA and NOMA are provided next.  

\begin{figure}[t]
\centering
\includegraphics[width=9cm]{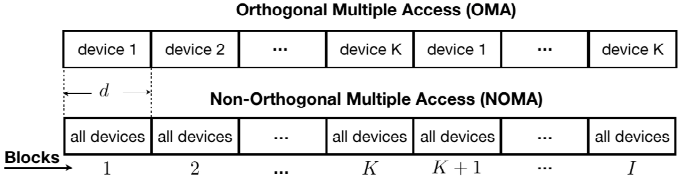}
\caption{Illustration of the transmission schedule for the considered multiple access protocols.}
\label{Fig: access}
\end{figure}

\subsubsection{ Orthogonal Multiple Access (OMA)} For orthogonal access, all devices  time-share the channel uses via {Time Division Multiple Access} (TDMA).  As illustrated in Fig.~\ref{Fig: access}, devices are scheduled successively in orthogonal {\color{black} blocks, and we assume that the total number of blocks satisfies $I=KT$, so that $T$ global gradient descent iterations \eqref{eq: Model update}  are implemented.}
In the  $i$-th block, with $i=K(t-1)+k$,  device $k$ transmits gradient information corresponding to the $t$-th iteration.  The signal received at the edge server during the $i$-th block is 
 \begin{align} \label{eq: recv signal}
\by^{[i]}_k= h_k^{[i]} \bx_k^{[i]} +\bz^{[i]}_k,
\end{align}
where $h_k^{[i]}\geq0$ is the channel gain for device $k$ in block $i$,
$\bx_k^{[i]}\in\mathbb{R}^d$ is an uncoded  function of the local gradient $\nabla {F}_k(\bw^{(t)})$, and $\bz_k^{[i]}$ is channel noise i.i.d. according to distribution $\mathcal{N}(0,{N_0}\bI)$.  We define as $\by^{(t)}=[\by^{[K(t-1)+1]},\cdots,\by^{[Kt]}]$ the vector collecting all signals received for iteration $t$ across $K$ blocks. 

\subsubsection{ Non-Orthogonal Multiple Access (NOMA)}  {\color{black}  For non-orthogonal access, we assume symbol-level synchronization among the devices that transmit simultaneously in each block. This can be achieved by using standard protocols such as the timing advance procedure in LTE and 5G NR \cite{mahmood2019time}.}
  In the $i$-th block, all devices upload the local gradients corresponding to the $t=i$-th iteration, {\color{black}and we have $I=T$ so that the number of blocks equals the number of iterations. }  The corresponding received signal is    
 \begin{align} \label{eq: power constraint}
\by^{[i]}= \sum_{k=1}^K h_k^{[i]} \bx_k^{[i]} +\bz^{[i]},
\end{align}
where $h_k^{[i]}$ and $\bz^{[i]}$ are defined as above; and signal $\bx_k^{[i]}\in \mathbb{R}^d$ encodes information about the local gradient $\nabla {F}_k(\bw^{(t)})$ with $t=i$.  For NOMA, we will also write $\by^{(t)}=\by^{[t]}$.

Note that for both forms of access, the transmit power constraint of a device is given as
\begin{equation}\label{eq: power constraint}
{(\text {Power constraint})} \quad \E[\|\bx_k^{[i]}\|^2] \leq P.
\end{equation}
{\color{black}
Accordingly, we define the maximum signal to noise ratio (SNR) as
\begin{equation}\label{eq: SNR}
\SNR_{\sf max}=\frac{P}{dN_0},
\end{equation}
where $dN_0$ represents the power of the channel nosies within one communication block. 
 We refer to  \eqref{eq: SNR} as the maximum SNR since devices may optimally transmit with a power strictly smaller than $P$ in \eqref{eq: power constraint} in order to satisfy the DP constraints.
}

\subsection{Differential Privacy }\label{sec: def DP}
As a threat model, we assume a ``honest-but-curious" edge server that may attempt to infer information about local data sets from the signals $\{\by^{(t)}\}_{t=1}^T$ received across $T$ successive iterations.  Note that, as discussed,  $T$ iterations correspond to $T$ communication blocks for NOMA and $TK$ blocks for OMA.  
The standard definition DP imposes a point-wise upper bound on the divergence between the distributions $P(\by|\mathcal{D})$ and $P(\by|\mathcal{D}')$
of the received signals $\by=\{\by^{(t)}\}_{t=1}^T$ conditioned on the use of either one of two ``neighboring" global data sets $\mathcal{D}$ and $\mathcal{D}'$.  The two neighboring data sets $\mathcal{D}$ and $\mathcal{D}'$ differ only by one sample at one of the devices. Defining the cardinality of the set difference for two sets $\mathcal{A}$ and $\mathcal{B}$ as $\|\mathcal{A}-\mathcal{B}\|_1$, we have the following formal definition.

\begin{definition}[Differential Privacy \cite{dwork2014algorithmic}] \emph{ The communication and learning protocol is $(\epsilon,\delta)$-differentially private, where $\epsilon>0$, and $\delta\in[0,1)$, if any two possible adjacent global datasets  $\mathcal{D}'=\bigcup_{k=1}^K \mathcal{D}'_k$ and $\mathcal{D}''=\bigcup_{k=1}^K \mathcal{D}_k''  $, with $\l\|\mathcal{D}'_j-\mathcal{D}''_j\r\|_1=1 $ for some device $j$ and $\l\|\mathcal{D}'_k-\mathcal{D}''_k\r\|_1=0$ for all $k\neq j$,
 we have the inequality 
\begin{equation}\label{eq: def dp}
P (\by |\mathcal{D}' ) \leq \exp(\epsilon)P(\by |\mathcal{D}'' )+\delta.
\end{equation}
 }
\end{definition}

The bound \eqref{eq: def dp} can be interpreted in terms of the test variable.
\begin{equation}\label{def: privacy loss}
{(\text{Differential privacy loss})}\quad \mathcal{L}_{\mathcal{D}',\mathcal{D}''}(\by )=\ln \frac{P(\by |\mathcal{D})}{P ( \by |\mathcal{D}')},
\end{equation} 
which is referred to as differential privacy loss. This corresponds to the log-likelihood ratio for the detection of neighboring data sets $\mathcal{D}'$ and $\mathcal{D}''$. The $(\epsilon,\delta)$-DP condition \eqref{eq: def dp} ensures that, for all possible adjacent global datasets $\mathcal{D}'$ and $\mathcal{D}''$, the absolute value of privacy loss variable \eqref{def: privacy loss} is bounded by $\epsilon$ with probability at least $1-\delta$, i.e., $\Pr(|\mathcal{L}_{\mathcal{D}',\mathcal{D}''}(\by)|\leq \epsilon)\geq 1-\delta$ (see Lemma 3.17 in \cite{dwork2014algorithmic}).
If $\epsilon$ and $\delta$ are suitably small, this makes it statistically impossible, even for an adversary that knows all data points in $\mathcal{D}$ except one,  to identify the remaining individual sample.

\subsection{Assumptions On the Loss Functions}
Finally, we list several standard assumptions we make on the loss functions and on its gradients.

\begin{assumption}[Smoothness]\label{assumption: Lipschitz}\emph{The global loss function $F(\bw)$ is smooth with constant $L>0$, that is, it is continuously differentiable and the gradient $\nabla F(\bw)$ is Lipschitz continuous with constant $L$, i.e., 
\begin{equation}\label{eq: Lipschitz}
\l\|\nabla F(\bw)-\nabla F(\bw')\r\|\leq L \l\|\bw-\bw'\r\|,  \text{for all } \bw,\bw' \in \mathbb{R}^d.
\end{equation}
Inequality \eqref{eq: Lipschitz} implies the following inequality
\begin{align}\label{eq: Lipschitz 2}
 F(\bw')\leq F(\bw)+ \nabla F(\bw)^{\sf T}&(\bw'-\bw)+\frac{L}{2}\l\|\bw-\bw'\r\|^2, \nn\\
&  \text{for all }  \bw,\bw' \in \mathbb{R}^d.
 \end{align}
}
\end{assumption}

\begin{assumption}[Polyak-Lojasiewicz Inequality]\label{assumption: PL}\emph{The optimization problem \eqref{eq: LP} has a non-empty solution set. Furthermore, denoting as $F^*$ the corresponding optimal function value, the global loss function 
$F(\bw)$ satisfies the Polyak-Lojasiewicz (PL) condition, that is, the following inequality holds for some constant $\mu>0$ 
 \begin{equation}\label{eq: PL ineq}
 \frac{1}{2}\l\|\nabla F(\bw)\r\|^2\geq\mu\l[F(\bw)-F^*\r].
 \end{equation}
 }
\end{assumption}
 
 The PL condition \eqref{eq: PL ineq} is significantly more general than the standard assumption of strong convexity \cite{karimi2016linear}. 
 A strong convex with constant $\mu>0$ implies the PL inequality with  same parameter $\mu$ \cite{bottou2018optimization}.  Note also that for a convex loss function $f(\cdot;\bu,v)$, e.g., for least squares and logistic regression, the strong convexity of function $F(\bw)$ follows from the addition of the regularizing term with $\lambda>0$.

%


\section{Orthogonal Multiple Access }\label{sec: OMA}

In this section, we consider the design and analysis of orthogonal multiple access with uncoded transmission and adaptive power control. To start, we assume that, at each iteration $t$, device $k$ transmits a scaled and noisy version of the gradient $\bx_k^{[i]}=\bx_k^{(t)}$ in block $i=K(t-1)+k$ as 
\begin{align}\label{eq: trans signal}
\bx_k^{(t)}=\alpha_k^{(t)}\l(D_k \nabla F_k\l(\bw^{(t)}\r) +{\bf n}_k^{(t)}\r).
\end{align}
In \eqref{eq: trans signal}, the artificial noise term ${\bf n}_k^{(t)}\sim\mathcal{N}(0,(\sigma_k^{(t)})^2\bI )$ is added in accordance to the standard Gaussian mechanism in the DP literature; and $\alpha_k^{(t)}\geq0$ is a scaling factor.  We note that, by \eqref{eq: trans signal}, the effective noise in the received signal \eqref{eq: recv signal} is given by the summation of channel and artificial noise.  The standard deviation of the effective noise is 
\begin{equation}\label{eq: effective noise}
m_k^{(t)}=\sqrt{(h_k^{(t)}\alpha_k^{(t)}\sigma_k^{(t)})^2+N_0}.
\end{equation}
 We are interested in optimizing over the sequences of parameters $(\alpha_k^{(1)} \!, \! \cdots \!, \!\alpha_k^{(T)})$ and $( \sigma_k^{(1)} \!, \! \cdots \! ,\!  \sigma_k^{(T)} )$ in order to maximize the learning performance under the $(\epsilon, \delta)$-DP constraint.  To this end, the remainder of this section first provides DP and convergence analysis, based on which the optimization problem is then formulated and solved.  
Throughout this section, we use the notation $h_k^{[i]}=h_k^{(t)}$, $\by_k^{[i]}=\by_k^{(t)}$ and $\bz_k^{[i]}=\bz_k^{(t)}$ for $i=K(t-1)+k$, and we make the following common assumption (see, e.g., \cite{friedlander2012hybrid,wu2019value,zhao2015stochastic}).  
\begin{assumption}[Bounded Sample-Wise Gradient]\label{ass: bounded sw gradient}
\emph{At any iteration $t$, for any training sample $(\bu,v)$, the gradient is upper bounded by a given constant $\gamma^{(t)}$, i.e., for all possible $(\bu,v)$ (not limited to those in data sets $\{\mathcal{D}_k\}$) we have the inequality 
\begin{align}
\|\nabla f(\bw^{(t)};\bu,v)\|\leq \gamma^{(t)}.
\end{align}}
\end{assumption}

\subsection{Differential Privacy Analysis}\label{sec: OMA DP analysis}
By standard results on DP,  the privacy level $(\epsilon,\delta)$ depends on the sensitivity of the function being disclosed, excluding the effect of noise, to the input data set.  More specifically, the sensitivity measures the amount by which a single individual data point can change the disclosed function in the worst case.  
For each device $k$, the edge server is assumed to be informed about parameters $\{\alpha_k^{(t)}\}$.  We assume here that those parameters are fixed constants that do not reveal information about the local datasets.
Hence, the only function of the data being disclosed is the received signal $\by_k^{(t)}$, upon subtraction of the effective noise.  The sensitivity $\Delta_k^{(t)}$ of the noiseless received signal $\by_k^{(t)}-\bz_k^{(t)}-h_k^{(t)}\alpha_k^{(t)}\bn_k^{(t)}$ is defined as
 \begin{align}\label{eq: def sensitivity}
(&\text{Sensitivity in OMA}) \
\Delta_k^{(t)} \! =\!\!\max_{\mathcal{D}'_k,\mathcal{D}''_k} \bigg\| {h_k^{(t)}\alpha_k^{(t)}}\times \nn\\
&\Big(\!\!\! \sum_{(\bu,v)\in\mathcal{D}'_k}\!\!\! \!\nabla f\big(\bw^{(t)};\bu,v\big)-\! \!\! \! \!\sum_{(\bu,v)\in\mathcal{D}''_k}\!\!\! \! \nabla f\big(\bw^{(t)};\bu,v\big)\Big)  \bigg\| , \!\!
\end{align}
where data sets $\mathcal{D}'_k$ and $\mathcal{D}''_k$ satisfy $\|\mathcal{D}'_k-\mathcal{D}''_k\|_{1}=1$.  By the triangular inequality and Assumption~\ref{ass: bounded sw gradient}, we have the bound 
\begin{align}\label{eq: bound def sensitivity}
\Delta_k^{(t)} \leq {2 h_k^{(t)}\alpha_k^{(t)}} \gamma^{(t)}.
\end{align}


 \begin{lemma}[Differential Privacy Guarantees for OMA]\label{lemma: privacy constraint}\emph{For any fixed sequence of parameters $\{\alpha_k^{(t)},\sigma_k^{(t)}\}$, federated gradient averaging via OMA  guarantees $(\epsilon,\delta)$-DP if the following condition is satisfied \vspace{-5pt}
\begin{align}
\sum_{t=1}^T\l(\frac{\sqrt{2}h_k^{(t)}\alpha_k^{(t)}\gamma^{(t)}}{m_k^{(t)}}\r)^2&\!\!\!\!\leq \l(\sqrt{\epsilon+\l[\mathcal{C}^{-1}\l({1}/{\delta}\r)\r]^2}-\mathcal{C}^{-1}\l({1}/{\delta}\r)\r)^2  \label{eq: privacy constraint} \\
&\overset{\Delta}{=}\mathcal{R}_{\sf dp}(\epsilon,\delta), \ \text{for all } k, \label{eq: def R_dp}
\end{align}
where $m_k^{(t)}$ in \eqref{eq: effective noise} is the standard deviation of the effective noise, and $\mathcal{C}^{-1}(x)$ is the inverse function of $\mathcal{C}(x)=\sqrt{\pi}xe^{x^2}$.
}\vspace{-5pt}
\end{lemma}\vspace{-5pt}
\proof  The proof is based on the  advanced composition theorem  \cite[Theorem 3.20]{dwork2014algorithmic} and is detailed in Appendix~\ref{proof: DP constraint}.  

Lemma~\ref{lemma: privacy constraint}, along with \eqref{eq: bound def sensitivity}, indicate that the privacy level depends on the sum of the per-iteration ratios $({\sqrt{2}h_k^{(t)}\alpha_k^{(t)}\gamma^{(t)}}/{m_k^{(t)}})^2$, which, by \eqref{eq: trans signal}, depend on the ratio between useful signal and effective noise powers. The effective noise level $m_k^{(t)}$ contributing to the privacy of device $k$ equals the sum of the channel noise power and of the noise added by device $k$ in \eqref{eq: trans signal}.  
The constraint \eqref{eq: privacy constraint} suggests that the effective noise variance can be adapted to the sequence of channel gains, as long as the impact on convergence is suitably accounted for.

\vspace{-10pt}
\subsection{Convergence Analysis}
At the $t$-th iteration, encompassing the blocks $i=K(t-1)+1,\cdots, Kt$, the edge server estimates the scaled local gradient $D_k\nabla F_k(\bw^{(t)})$ as $(h_k^{(t)}\alpha_k^{(t)})^{-1} \by_k^{(t)}$, and then the global gradient is estimated as
\begin{align}\vspace{-10pt}
\widehat{\nabla F}(\bw^{(t)})&=\frac{1}{D_{\sf tot}}\sum_{k=1}^K\l(h_k^{(t)}\alpha_k^{(t)}\r)^{-1} \by_k^{(t)} \nn\\
&=\frac{1}{D_{\sf tot}}\sum_{k=1}^K D_k\nabla F_k\l(\bw^{(t)}\r) \!+\! {\bf n}_k^{\l(t\r)} \!\!+ \!\!\l(h_k^{(t)}\alpha_k^{(t)}\r)^{-1}\!\!\bz_k^{(t)}.\vspace{-5pt}
\end{align}
Building on standard results on gradient descent with noisy gradient \cite{bottou2018optimization}, we have the following bound on the average optimality gap at the end of iteration $T$.
 \begin{lemma}[Optimality Gap Bound for OMA] \label{lemma: convergence}\emph{Under Assumptions~\ref{assumption: Lipschitz} and \ref{assumption: PL},
 for a learning rate $\eta=1/L$, after $T$ iterations the average optimality gap is  upper bounded as
\begin{align} \label{eq: result convergence}
 \E&\l[F\l(\bw^{(T+1)}\r)-F^*\r]
 \leq \l(1-\frac{\mu}{L}\r)^{T}\l[F\l(\bw^{(1)}\r)-F^*\r] \nn\\
 &\qquad +\frac{d}{2LD_{\sf tot}^2}\sum_{t=1}^T \l(1-\frac{\mu}{L}\r)^{T-t}\sum_{k=1}^K \bigg(\frac{m_k^{(t)}}{h_k^{(t)}\alpha_k^{(t)}}\bigg)^2 , 
 \end{align}
where the standard deviation $m_k^{(t)}$ of the effective noise is defined in \eqref{eq: effective noise}. }
\end{lemma}
  \vspace{-5pt}
\proof See Appendix~\ref{proof: convergence}.

The first term in \eqref{eq: result convergence} reflects the standard geometric decay of the initial optimality gap ($F(\bw^{(1)})-F^*$) as $T$ increases, while the second accounts 
for the impact of the effective additive noise powers  \eqref{eq: privacy constraint}.  Interestingly, the bound \eqref{eq: result convergence} suggests that noise added in the initial iterations is less damaging to the final optimality gap than the noise added in later iterations.  This is because the contribution of the noise added at iteration $t$ is discounted by a factor $(1-\mu/L)^{T-t}$.  We will leverage this result in the next section to optimize power allocation. 

\subsection{Optimization}\label{sec: OMA optimization}
In this section, we are interested in minimizing the optimality bound in Lemma~\ref{lemma: convergence} under  $(\epsilon, \delta)$-DP constraint \eqref{eq: privacy constraint} and the power  constraints \eqref{eq: power constraint}, for all $K$ devices across $T$ iterations. {\color{black} Note that, for the objective function \eqref{eq: result convergence}, the optimization variables only exist in the second term. By replacing $m_k^{(t)}$ with its definition given in \eqref{eq: effective noise}, the resulting optimization problem ({\bf OMA Opt.}) of interest   is formulated as }
 \begin{subequations}\label{eq: OMA OPT}
\begin{align}
 \min_{{\{\sigma_k^{(t)},\alpha_k^{(t)}\}}_{k=1}^K}
\sum_{t=1}^{T}\l(1-\frac{\mu}{L}\r)^{-t} \!\sum_{k=1}^K \!\! \bigg[(\sigma_k^{(t)})^2\!\!+\!&\bigg(\frac{\sqrt{N_0}}{h_k^{(t)}\alpha_k^{(t)}}\bigg)^2\bigg] \!\!\label{eq: OMA OPTa}\\
{\rm s.t.} \   \sum_{t=1}^{T}\frac{(\sqrt{2}\gamma^{(t)})^2}{(\sigma_k^{(t)})^2+ {N_0} /(h_k^{(t)}\alpha_k^{(t)})^{2}}\leq &\mathcal{R}_{\sf dp}(\epsilon,\delta), \ \forall k,\!\! \label{eq: OMA OPTb}\\
 (\alpha_k^{(t)})^2\l[(D_kG_k^{(t)})^2+d(\sigma_k^{(t)})^2\r] \leq &P , \quad \forall k,t, \label{eq: OMA OPTc}
 \end{align}
 \end{subequations}
 where $\mathcal{R}_{\sf dp}(\epsilon,\delta)$ is defined in \eqref{eq: def R_dp}, and {\color{black}Parameter $G_k^{(t)}$ represents an upper bound on the norm of the local gradient as $ \l\|\nabla F_k\l(\bw^{(t)}\r)\r\|\leq G_k^{(t)}$. By Assumption~\ref {ass: bounded sw gradient}, we have $G_k^{(t)}\leq \gamma^{(t)}$.}
 Under OMA, the optimization \eqref{eq: OMA OPTa}-\eqref{eq: OMA OPTc} over the additive noise deviations $\{\sigma_k^{(1)}, \cdots, \sigma_k^{(T)}\}$ and scaling factors $\{\alpha_k^{(1)}, \cdots, \alpha_k^{(T)}\}$ for each devices $k$ can be carried out in parallel.  The corresponding problem ({\bf OMA Local Opt.}) to be solved by device $k$ is
 \begin{subequations}\label{eq: OMA Local OPT}
\begin{align}
\min_{{\{\sigma_k^{(t)},\alpha_k^{(t)}\}}}&\quad 
\sum_{t=1}^{T} \l(1-\frac{\mu}{L}\r)^{-t} \bigg[(\sigma_k^{(t)})^2+\bigg(\frac{\sqrt{N_0}}{h_k^{(t)}\alpha_k^{(t)}}\bigg)^2\bigg] \label{eq: OMA Local OPTa}\\
{\rm s.t.} \quad &\sum_{t=1}^{T}\frac{(\sqrt{2}\gamma^{(t)})^2}{(\sigma_k^{(t)})^2+ {N_0} /(h_k^{(t)}\alpha_k^{(t)})^{2}}\leq \mathcal{R}_{\sf dp}(\epsilon,\delta), \label{eq: OMA Local OPTb}\\
&(\alpha_k^{(t)})^2\l[(D_kG_k^{(t)})^2+d(\sigma_k^{(t)})^2\r]  \leq P , \quad \forall  t. \label{eq: OMA Local OPTc}
 \end{align}
  \end{subequations}
  
Without the DP constraint \eqref{eq: OMA Local OPTb},  the optimal solution to problem \eqref{eq: OMA Local OPT} is to fully use the power budget $P$ for the transmission of the local gradient, i.e., to set $\alpha_k^{(t)}=\sqrt{P}/(D_kG_k^{(t)})$ and $\sigma_k^{(t)}=0$ for all $k$ and $t$.  Due to the DP constraint, we now show that this may not be the optimal solution if the privacy condition is sufficiently strict.  

Before detailing offline and online solutions, it is useful to observe that, in order for constraint \eqref{eq: OMA Local OPTb} to guarantee $(\epsilon,\delta)$-DP, by leave $t$, it is necessary that the parameters $G_k^{(t)}$ be fixed at each iteration $t$ in a way that does not depend on the local data sets. We will return to this point when discussing online methods.

 \subsubsection{Offline Optimization}
 We first assume that the parameters $\{h_k^{(t)}, \gamma^{(t)}, G_k^{(t)}\}$ are known beforehand so that problem \eqref{eq: OMA Local OPTa}-\eqref{eq: OMA Local OPTc} can be tackled offline.  
 As we show in Appendix~\ref{proof: optimal solution TDMA}, problem \eqref{eq: OMA Local OPTa}-\eqref{eq: OMA Local OPTc} can be converted into a convex program via a change of variables. The resulting optimal solution is described in the following theorem. 
 
\begin{theorem}\label{thm: optimal solution TDMA} \emph{The optimal offline solution of problem \eqref{eq: OMA OPTa}-\eqref{eq: OMA OPTc}  under OMA is given as follows: 
\begin{itemize}
\item If condition 
\begin{equation}\label{eq: free dp condition}
\sum_{t=1}^{T} \frac{P(\sqrt{2}\gamma^{(t)} h_k^{(t)})^2}{{N_0}(D_kG_k^{(t)})^2}<\mathcal{R}_{\sf dp}(\epsilon,\delta)
\end{equation}
holds, there exists a unique optimal solution given as $(\alpha_k^{(t)})_{\sf opt}=\sqrt{P}/(D_kG_k^{(t)})$ and  $(\sigma_k^{(t)})_{\sf opt}=0$. In this case, the power budget $P$ is fully used for the transmission of the local gradient, and the channel noise is sufficient to guarantee privacy.  The optimal solution is identical as that of without DP constraint,  and privacy is hence obtained ``for free";
\item Otherwise, there exist multiple optimal solutions, and the solution that minimizes the transmit power is
\begin{align}
 (\alpha_k^{(t)})_{\sf opt}&= \min\bigg\{   \frac{\sqrt{N_0}  (2{\zeta_k})^{-\frac{1}{4}}}{h_k^{(t)} \sqrt{\gamma^{(t)}} }\l(1-\frac{\mu}{L}\r)^{-{t}/{4}}\!\!\!\!, \frac{\sqrt{ P}}{D_kG_k^{(t)}}\bigg\} \label{eq: solution alpha local}\\
 (\sigma_k^{(t)})_{\sf opt}&= 0,  
 \end{align}
where the value of parameter $\zeta_k$ can be obtained by bisection to satisfy the constraint 
\begin{align}
\sum_{t=1}^{T} {(\sqrt{2}\gamma^{(t)})^2} \min&\bigg\{\frac{(1-\mu/L)^{-t/2}}{\sqrt{2\zeta_k}\gamma^{(t)}}, \frac{P (h_k^{(t)})^2}{(\sqrt{N_0}D_kG_k^{(t)})^2}\bigg\}\nn\\
&\qquad \qquad \quad \qquad= \mathcal{R}_{\sf dp}(\epsilon,\delta).
\end{align}
 In this case, the transmitted power needs to be scaled down in order to leverage the channel noise to ensure $(\epsilon,\delta)$-DP.
\end{itemize}
}
\end{theorem} \vspace{-5pt}
\proof The proof is detailed in Appendix~\ref{proof: optimal solution TDMA}.

A first interesting observation from Theorem~\ref{thm: optimal solution TDMA} is that it is optimal for the devices not to add noise to the transmitted signals \eqref{eq: trans signal}.  It is, in fact, sufficient to scale down their transmitted powers, via the choice of $\alpha_k^{(t)}$, with smaller powers transmitted when more stringent DP constraints are imposed. 
Second, when condition \eqref{eq: free dp condition} is satisfied, privacy is obtained ``for free", that is, without affecting the learning performance of the system, as the devices can use their full power.  
Third, condition \eqref{eq: free dp condition} is less strict as $D_k$ increases, showing that  devices with larger datasets can attain privacy ``for free" over a broader range of SNR levels. 
Finally, we note that it is generally suboptimal to use a time-invariant policy that sets the scaling factor $\alpha_k^{(t)}$ as a constant.

\subsubsection{Online Optimization}\label{sec: OMA online}
Theorem~\ref{thm: optimal solution TDMA} assumes that the sequence of parameters $\{h_k^{(t)}, \gamma^{(t)}, G_k^{(t)}\}$ is known a priori so as to enable offline optimization.  Here we describe a heuristic online approach that leverages iterative one-step-ahead optimization based on predicted values for the future parameters $\{h_k^{(t)}, \gamma^{(t)}, G_k^{(t)}\}$.

To elaborate, assume that, at each iteration $t$, we have predicted values $\{\widehat{h}_k^{(t')}, \widehat{\gamma}^{(t')}, \widehat{G}_k^{(t')}\}$ for $t'=t,t+1,\!\cdots\!, T$ and that the accumulated DP loss is given by  $\!\mathcal{L}_k^{(t-1)}\!\!\!=\!\!\sum_{t'=1}^{t-1}(\!{\sqrt{2}h_k^{(t')}\alpha_k^{(t')}\gamma^{(t')}}/{m_k^{(t')}})^2$ from \eqref{eq: privacy constraint}.  As summarized in Algorithm 1, we propose to apply the solution in Theorem 1 to the interval $(t,t+1,\cdots, T)$ by replacing the true parameters  $\{h_k^{(t)}, \gamma^{(t)}, G_k^{(t)}\}$ with the estimates  $\{\widehat{h}_k^{(t)}, \widehat{\gamma}^{(t)}, \widehat{G}_k^{(t)}\}$ and the DP constraint with the residual $\mathcal{R}_{\sf dp}(\epsilon,\delta)-\mathcal{L}_k^{(t-1)}$.  The produced scaling factors $\alpha_k^{(t)}$ are then applied, and the procedure is repeated for iteration $t+1$.  We now discuss the problem of prediction of parameters  $\{{h}_k^{(t)}, {\gamma}^{(t)}, {G}_k^{(t)}\}$.

To start, we model the sequence of fading channels $\{g_k^{[i]}\}$ via an autoregressive (AR) Rician model.  We note that the method can be directly extended to other probabilistic models.  Accordingly, each channel gain $h_k^{[i]}$ is obtained as $h_k^{[i]}=|g_k^{[i]}|$, where the complex channel coefficient  $\{g_k^{[i]}\}$ is given as  
\begin{align}\label{eq:channel model}
g_k^{[i]}=\sqrt{\frac{\kappa_k}{\kappa_k+1}}+\sqrt{\frac{1}{\kappa_k+1}} r_k^{[i]},
\end{align}
with $\kappa_k$ being the Rice parameter and the stochastic diffuse component $r_k^{[i]}$ following an $AR(1)$ process.  We specifically write  $r_k^{[i+1]}=\rho_k r_k^{[i]}+ \sqrt{1-\rho_k^2} \widetilde{r}_k^{[i]}$, with temporal correlation coefficient $0\leq\rho_k\leq1$, and $\widetilde{r}_k^{[i]}\sim \mathcal {CN}(0,1)$ being an i.i.d. innovation process.  Given the current CSI $g_k^{[i]}$, the future channel power $(h_k^{[j]})^2=|g_k^{[i]}|^2$ for $j>i$ can be predicted via minimum mean squared error (MMSE) estimation as 
\begin{align}\label{eq: future channels est}
(\widehat{h}_k^{[j]})^2=\E\big[(h_k^{[j]})^2 \big| g_k^{[i]}\big]={\frac{\kappa_k+(\rho_k^{j-i})^{2}}{\kappa_k+1}}|g_k^{[i]}|^2+\frac{1-(\rho_k^{j-i})^2}{\kappa_k+1}.
\end{align}

Next, we discuss the estimations of parameters $\{\gamma^{(t)}\}$ and $\{G_k^{(t)}\}$.  {\color{black}Parameter $\gamma^{(t)}$ is by definition independent of the local data sets, and is typically determined by clipping the local gradient before transmission to the server \cite{abadi2016deep,chen2020understanding}.} To this end, in \eqref{eq: local gradient}, we substitute the per-sample gradient $\nabla f\l(\bw^{(t)};\bu,v\r)$ with its clipped version
\begin{align}\label{eq: clipped local update}
&{(\text {Clipped per-sample gradient})}  \nn\\
 &\overline{\nabla f}\l(\bw^{(t)};\bu,v\r)= \min \bigg\{1, \frac{\widehat{\gamma}}{\|\nabla f\l(\bw^{(t)};\bu,v\r)\|}\bigg\}\nn\\
& \qquad \qquad \qquad \qquad \qquad \qquad \qquad \times\nabla f\big(\bw^{(t)};\bu,v\big)
\end{align}
for some fixed threshold $\widehat{\gamma}>0$.

The definition of the parameters $\{G_k^{(t)}\}$
makes them generally data-dependent.  In order to avoid leaking additional information about the data to the server, we propose to predict bounds $\{\widehat{G}_k^{(t')}\}$ for $t' \geq t$ based on an additional signal broadcast by the server.  Specifically, we let the edge server transmit the positive scalar $\|\by_k^{(t-1)}\|/(h_k^{(t-1)}\alpha_k^{(t-1)})$ back to device $k$ in addition to the broadcast signal $\bw^{(t)}$.  Basing the predictions $\widehat{G}_k^{(t')}$ on the past received signal $\by_k^{(t-1)}$ does not affect privacy, since the privacy loss due to the reception of $\by_k^{(t-1)}$ at the edge server is accounted for by $\mathcal{L}_k^{(t-1)}$.
At each iteration $t$, any device $k$ sets
\begin{subequations}\label{eq: value G_k}
\begin{align} 
&\widehat{G}_k^{(t)}=\left\{\begin{matrix}
\|\by_k^{(t-1)}\|/(h_k^{(t-1)}\alpha_k^{(t-1)}D_k), &t>1,\\
\widehat{\gamma},&t=1. 
\end{matrix}\right. \label{eq: value G_k a}
\\
\text{and} \ &\widehat{G}_k^{(t')}=\widehat{G}_k^{(t)}, \forall \ t'>t. \label{eq: value G_k b}\vspace{-10pt}
\end{align}
\end{subequations}
Furthermore, in order to ensure constraint on the bounded local gradient $ \l\|\nabla F_k\l(\bw^{(t)}\r)\r\|\leq G_k^{(t)}$,
we clip the local gradient for transmission as 
\begin{align}\label{eq: clipped local update}
&{(\text {Clipped gradient transmission in OMA})} \nn\\
 &\overline{\nabla f}\l(\bw^{(t)};\bu,v\r)= \min \bigg\{1, \frac{\widehat{\gamma}}{\|\nabla f\l(\bw^{(t)};\bu,v\r)\|}\bigg\}\nn\\
& \qquad \qquad \qquad \qquad \qquad \qquad \qquad \times\nabla f\big(\bw^{(t)};\bu,v\big)
\end{align}
with $\overline{\nabla F}_k=\frac{1}{{D}_k}\sum_{(\bu,v)\in\mathcal{D}_k}  \overline{\nabla f}\l(\bw^{(t)};\bu,v\r)+\lambda \nabla R(\bw)$. 

Finally, we observe that, strictly speaking, the analysis of convergence in  Lemma~\ref{lemma: convergence} should be modified in order to account for clipping, but we found the heuristic approach summarized in Algorithm 1 to perform well in practice.

\begin{algorithm}
{\bf Input:} DP level $\mathcal{R}_{\sf dp}$, channel noise $\sqrt{N_0}$, channel correlation $\rho$, clipping threshold $\gamma^{(t)}=\widehat{\gamma}$ 

{\bf Initialize:} Local privacy loss $\mathcal{L}_k^{(0)}=0$

{\bf For each iteration:} $t=1,\dots,T$

\quad \quad {\bf For each device:} $k=1,\dots,K$

\quad\quad \quad Receive $\bw^{(t)}$ from edge server

\quad\quad \quad Update local model by \eqref{eq: clipped local update}

\quad\quad \quad  {\bf If $t>1$}

\quad\quad \quad \quad {Receive $\by_k^{(t-1)}/(h_k^{(t-1)}\alpha_k^{(t-1)})$} from edge server

\quad\quad \quad {\bf end}

\quad \quad \quad   {Compute predictors $\{\widehat{h}_k^{(t')},\widehat{G}_k^{(t')}\}$ via \eqref{eq: future channels est} and  \eqref{eq: value G_k} 

\quad \quad \quad \quad for $t'\in [t,\cdots,T]$ with $\widehat{h}_k^{(t)}={h}_k^{(t)}$ }

\quad \quad \quad  {Apply Theorem~\ref{thm: optimal solution TDMA} over the time interval $[t,\cdots,T]$ with 

\quad\quad \quad \quad $\{{h}_k^{(t')}\leftarrow\widehat{h}_k^{(t')},\ {G}_k^{(t')}\leftarrow\widehat{G}_k^{(t')},\ \gamma^{(t)}\leftarrow \widehat{\gamma} \}$ 

\quad\quad \quad \quad and residual DP constraint $\mathcal{R}_{\sf dp}-\mathcal{L}_k^{(t-1)}$}

  \quad\quad \quad  {Use optimized scaling factor  $\alpha_k^{(t)}$ to transmit \eqref{eq: clipped local update}}

  \quad\quad \quad  {Update local privacy loss as 
  
 \quad\quad \quad \quad  $\mathcal{L}_k^{(t)}=\mathcal{L}_k^{(t-1)}+\frac{(\sqrt{2} \gamma^{(t)}h_k^{(t)}\alpha_k^{(t)})^2}{N_0}$ }
 
\quad \quad {\bf end}

 {\bf end}

\caption{Online Scheme for OMA}
\label{algorithm: online orthogonal}
\end{algorithm}

 \section{Non-orthogonal Multiple Access}\label{sec: NOMA}

 In this section, we consider the design and analysis of NOMA. For the $t$-th iteration,  local gradients are transmitted using the uncoded strategy \eqref{eq: trans signal}.  As in \cite{zhu2019broadband,yang2020federated, amiri2020machine, sery2020analog,seif2020wireless}, we select the scaling factors $\alpha_k^{(t)}$ so as to ensure that, in the absence of noise, the edge server can recover a scaled version of the global gradient \eqref{eq: global gradient}.  Accordingly, we set 
 \begin{align}\label{eq: gradient alignment}
 \text{(Gradient Alignment)}\quad h_k^{(t)}\alpha_k^{(t)}=c^{(t)},
 \end{align}
 for some constant $c^{(t)}$. We note that the effective noise in NOMA is equal to the summation of the channel noise and the contributions of artificial noise from all devices, and its standard deviation is given by
\begin{align}\label{eq: noise in y NOMA}
m^{(t)}=\sqrt{(c^{(t)})^2\sum_{k=1}^K(\sigma_k^{(t)})^2+N_0}.
\end{align} 
As per \eqref{eq: gradient alignment}, in this section, we are optimizing  over the parameters $(c^{(1)},\cdots, c^{(T)})$ as well as over the added noise power $(\sigma_k^{(1)}, \cdots, \sigma_k^{(T)})$. Throughout this section, we denote $h_k^{[i]}=h_k^{(t)}$, and $\bz^{[i]}=\bz^{(t)}$ for $i=t$.

\subsection{Differential Privacy Analysis}
As discussed in Section~\ref{sec: OMA DP analysis}, the DP level depends on the sensitivity of the function being disclosed, which, in NOMA, for the same reasons discussed in Section~\ref{sec: OMA DP analysis}, is the received noiseless aggregated signal.  The sensitivity to change in the data set of device $k$ is accordingly defined as 
 \begin{align}
&{(\text{Sensitivity in NOMA})} \ \ \Delta_k^{(t)}=\max_{\mathcal{D}'_k,\mathcal{D}''_k} \bigg\| {c^{(t)}} \times\nn\\
& \qquad\Big(\!\!\! \sum_{(\bu,v)\in\mathcal{D}'}\!\!\!\!\nabla f\big(\bw^{(t)};\bu,v\big) \! - \!\!\!\!\!\sum_{(\bu,v)\in\mathcal{D}''}\!\!\!\! \nabla f\big(\bw^{(t)};\bu,v\big)\Big)   \bigg\|,
\end{align}
 where $\l\|\mathcal{D}'_k-\mathcal{D}''_k\r\|_1=1 $, $\l\|\mathcal{D}'_j-\mathcal{D}''_j\r\|_1=0 $ for all $j\neq k$, and $\mathcal{D}'=\bigcup_{k=1}^K \mathcal{D}_k' $, $\mathcal{D}''=\bigcup_{k=1}^K \mathcal{D}_k''  $.  By Assumption~\ref{ass: bounded sw gradient}, we can bound the sensitivity as 
 \begin{align}
 \Delta_k^{(t)}\leq {2 c^{(t)}} \gamma^{(t)}.
\end{align} 
Then, the DP guarantees for NOMA are given as follows. 
\begin{lemma}[Differential Privacy Guarantees for NOMA]\label{lemma: privacy constraint NOMA}\emph{Federated gradient averaging via NOMA  guarantees $(\epsilon,\delta)$-DP if the following condition is satisfied
\begin{align}
\sum_{t=1}^T\l(\frac{\sqrt{2}c^{(t)}\gamma^{(t)}}{m^{(t)}}\r)^2&\leq \mathcal{R}_{\sf dp}(\epsilon,\delta), \ \text{for all } k. \label{eq: DP constraint NOMA}
\end{align}
where $m^{(t)}$ is the standard deviation of the effective noise~\eqref{eq: noise in y NOMA}.
}
\end{lemma} \vspace{-5pt}
{\color{black}
\proof The proof follows in a manner similar to Lemma~\ref{lemma: privacy constraint} by replacing the sensitivity and effective noise with those defined in NOMA. }

Lemma~\ref{lemma: privacy constraint NOMA} indicates that the effective noise contributing to the privacy of each device $k$ is given by the sum of channel noise and the privacy-inducing noise added by all devices.  This is an important advantage of NOMA, which was also observed in \cite{seif2020wireless}.
%

 \subsection{Convergence Analysis}
At the $t$-th iteration, the edge server estimates the global gradient as 
\begin{align}
\widehat{\nabla F}(\bw^{(t)})&=\frac{1}{D_{\sf tot}}\l(c^{(t)}\r)^{-1} \by^{(t)} \nn\\
&=\frac{1}{D_{\sf tot}}\sum_{k=1}^K D_k\nabla F_k\big(\bw^{(t)}\big) +{\bf n}_k^{\l(t\r)} + \l(c^{(t)}\r)^{-1}\bz_k^{(t)}.
\end{align}

 \begin{lemma}[Optimality Gap Bound for NOMA] \label{lemma: convergence NOMA}\emph{Under Assumptions~\ref{assumption: Lipschitz} and \ref{assumption: PL},
 for a learning rate $\eta=1/L$, after $T$ iterations the average optimality gap is  upper bounded as
\begin{align} \label{eq: result convergence NOMA}
 &\E\l[F\l(\bw^{(T+1)}\r)-F^*\r]
 \leq \l(1-\frac{\mu}{L}\r)^{T}\l[F\big(\bw^{(1)}\big)-F^*\r] \nn\\
 &\qquad \qquad \qquad+\frac{d}{2LD_{\sf tot}^2}\sum_{t=1}^T \l(1-\frac{\mu}{L}\r)^{T-t}\bigg(\frac{m^{(t)}}{c^{(t)}}\bigg)^2 , 
 \end{align}
where the standard deviation $m^{(t)}$ of the effective noise is defined in \eqref{eq: noise in y NOMA}.}
\end{lemma} \vspace{-5pt}
\proof
The proof follows via the same steps reported in Appendix~\ref{proof: convergence} by replacing the summation of \eqref{eq: effective noise} with \eqref{eq: noise in y NOMA}.

\subsection{Optimization}
In this section, we are interested in minimizing the optimality bound in Lemma~\ref{lemma: convergence NOMA} under the $(\epsilon, \delta)$-DP constraint \eqref{eq: DP constraint NOMA} and the power  constraints \eqref{eq: power constraint} across $T$ iterations.  The resulting optimization problem for NOMA  ({\bf NOMA Opt.}) is  formulated~as    
\begin{subequations}\label{NOMA OPT}
 \begin{align}
  &\min_{{\{\sigma_k^{(t)}, c^{(t)}\}}_{k=1}^K} \ 
\sum_{t=1}^{T} \l(1-\frac{\mu}{L}\r)^{-t}\sum_{k=1}^K(\sigma_k^{(t)})^2+ {N_0} /(c^{(t)})^{2} \!\!\!\label{NOMA OPTa}\\
&\qquad  {\rm s.t.} \ \sum_{t=1}^{T} \frac{(\sqrt{2}\gamma^{(t)})^2}{\sum_{k=1}^K(\sigma_k^{(t)})^2+ {N_0} /(c^{(t)})^{2}}\leq \mathcal{R}_{\sf dp}(\epsilon, \delta)  \label{NOMA OPTb}\\
 &\qquad \quad \l(\frac{c^{(t)}}{h_k^{(t)}}\r)^2\l[(D_kG_k^{(t)})^2+d(\sigma_k^{(t)})^2\r] \leq P , \ \forall k,t.   \label{NOMA OPTc}
 \end{align}
 \end{subequations}

 Without the DP constraint \eqref{NOMA OPTb}, the optimal solution to problem \eqref{NOMA OPT}  is determined by the devices with the smallest value of the ratio ${h_k^{(t)}}/{(D_kG_k^{(t)})}$ due to the need to satisfy the gradient alignment condition \eqref{eq: gradient alignment}.  In particular, the optimal solution prescribes that such devices use the full power budget $P$ to transmit the local gradient while the other devices transmit at the maximum power allowed under condition \eqref{eq: gradient alignment}, i.e., $c^{(t)}=\sqrt{P}\min_kh_k^{(t)}/(D_kG_k^{(t)})$ and $\sigma_k^{(t)}=0$.  We will see next that this is no longer the  optimal solution under sufficiently strict DP constraints.  
 
 \subsubsection{Offline Optimization}
 We first assume that the parameters $\{h_k^{(t)}, \gamma^{(t)}, G_k^{(t)}\}$ are known beforehand so that problem \eqref{NOMA OPTa}-\eqref{NOMA OPTc} can be tackled offline.  With a change of the variables, the problem can be shown to be convex.  
Unlike the previously studied problem for OMA, the optimization \eqref{NOMA OPTa}-\eqref{NOMA OPTc} cannot be solved in parallel across the devices. 
\begin{theorem}\label{thm: optimal solution non-orth}\emph{The optimal offline solution of problem \eqref{NOMA OPTa}-\eqref{NOMA OPTc} under NOMA is given as follows: 
\begin{itemize}
\item If condition 
\begin{align}
\frac{2P}{N_0}\sum_{t=1}^{T} (\gamma^{(t)})^2 \min_k \l(\frac{h_k^{(t)}}{D_kG_k^{(t)}}\r)^2<\mathcal{R}_{\sf dp}(\epsilon,\delta) \label{eq: condition free DP NOMA}
\end{align}
holds, there exists a unique optimal solution given as $(c^{(t)})_{\sf opt}=\sqrt{P}\min_k {h_k^{(t)}}/{(D_kG_k^{(t)})}$ and  $(\sigma_k^{(t)})_{\sf opt}=0$. In this case, the devices with smallest value of the ratio ${h_k^{(t)}}/{(D_kG_k^{(t)})}$ transmit using the full power budget $P$, while the other devices do not use full power. Therefore, under the gradient alignment condition \eqref{eq: gradient alignment}, privacy is obtained ``for free";
\item Otherwise, there exist multiple optimal solutions, and the solution that minimizes the transmit power at all devices~is
\begin{align}
 (c^{(t)})_{\sf opt}&= \min\bigg\{   \frac{\sqrt{N_0}  (2{\zeta})^{-\frac{1}{4}}}{\sqrt{\gamma^{(t)}} }\l(1-\frac{\mu}{L}\r)^{-{t}/{4}}, \nn\\
 &\qquad \qquad \qquad\qquad\sqrt{ P} \min_k \frac{h_k^{(t)}}{D_kG_k^{(t)}}\bigg\} \\
 (\sigma_k^{(t)})_{\sf opt}&= 0, 
  \end{align}
where the value of parameter $\zeta$ can be obtained by bisection to satisfy the constraint 
\begin{align}
\!\!\!\sum_{t=1}^{T} {(\sqrt{2}\gamma^{(t)})^2} \min\!\bigg\{\frac{(1-\mu/L)^{-t/2}}{\sqrt{2\zeta}\gamma^{(t)}},&\frac{P}{N_0} \min_k \bigg(\frac{ h_k^{(t)}}{D_kG_k^{(t)}}\bigg)^2\bigg\} \nn\\
&= \mathcal{R}_{\sf dp}(\epsilon,\delta). \label{eq: solution of lambda}
\end{align}
 In this case, all the transmitted powers need to be scaled down in order to leverage the channel noise to ensure $(\epsilon,\delta)$-DP.
\end{itemize}
}
\end{theorem} \vspace{-5pt}
{\color{black}
\proof The proof follows via the same steps of Theorem 1 by replacing the local optimization problem in OMA with the optimization problem of the device with smallest value of the ratio ${h_k^{(t)}}/{(D_kG_k^{(t)})}$.}

In a manner similar to OMA, Theorem~\ref{thm: optimal solution non-orth} demonstrates that it is optimal for devices not to add further noise to the transmitted signals, i.e., to set $\bn_k^{(t)}=0$ in \eqref{eq: trans signal}.  
Furthermore, under condition \eqref{eq: condition free DP NOMA}, privacy is attainable ``for free" since the optimal solution coincides with that obtained when excluding the DP constraint \eqref{NOMA OPTb}.  As for OMA, increasing the size $D_k$ of the data sets makes condition \eqref{eq: condition free DP NOMA} less restrictive.

\subsubsection{Online Optimization}  With the offline results in Theorem 2, we are ready to describe a heuristic online approach for NOMA which follows the same logic as in Section~\ref{sec: OMA optimization}.  In particular, at each iteration $t$, the edge server solves problem \eqref{NOMA OPT} over the interval $[t,\cdots, T]$ of current and future time instants by using estimated parameters $\{h_k^{(t)}, \gamma^{(t)}, G_k^{(t)}\}$, and imposing the residual DP constraint for each device.  We note that the optimization problem for NOMA is solved at edge server  with the known values of $\{D_k\}$.

To detail the procedure summarized in Algorithm 2, channels are predicted as in \eqref{eq: future channels est}.  Parameter $\widehat{\gamma}$ is set through the clipped per-sample gradient \eqref{eq: clipped local update}. Finally, estimates $\{\widehat{G}_k^{(t)}\}$ are obtained by using the received signal of the last iteration as described in OMA, but averaged with the number of global data set, which is given as 
\begin{subequations} \label{eq: est DkGk}
\begin{align}
& \widehat{G}_k^{(t)}=\left\{\begin{matrix}
\|\by^{(t-1)}\|/(c^{(t-1)}D_{\sf tot}), &t>1, \forall k,\\
\widehat{\gamma},&t=1, \forall k,
\end{matrix}\right. \label{eq: est DkGk a}
\\
& \widehat{G}_k^{(t')}=\widehat{G}_k^{(t)}, \forall \ t'>t, \forall k. \label{eq: est DkGk b}
\end{align}
\end{subequations}
One last issue to consider is that the optimized $c^{(t)}$ may violate the power constraint due to the use of estimated parameters. We hence modify the clipped gradient transmission as 
\begin{align}\label{eq: clipped transmission NOMA}
&{(\text {Clipped gradient transmission in NOMA})}  \nn\\
 & \bx_k^{(t)}= \min\l\{1, \frac{\sqrt{P}h_k^{(t)}}{c^{(t)}D_k\|\overline{\nabla F}_k \|}\r\}\frac{c^{(t)}}{h_k^{(t)}}D_k  \overline{\nabla F}_k \l(\bw^{(t)}\r).
\end{align}

As for NOMA, we make no claims of optimality, and we test the performance of the proposed online scheme via numerical results in the next section. 

\begin{algorithm}
{\bf Input:} DP level $\mathcal{R}_{\sf dp}$, channel noise $\sqrt{N_0}$, channel correlation $\rho$, clipping threshold $\gamma^{(t)}=\widehat{\gamma}$ 

{\bf Initialize:} Privacy loss $\mathcal{L}_g^{(0)}=0$.

{\bf For each iteration:} $t=1,\cdots, T$

\quad \quad {\bf For edge server:} 

\quad \quad \quad  {Compute predictors $\{\widehat{h}_k^{(t')},\widehat{G}_k^{(t')}\}$ via \eqref{eq: future channels est} and  \eqref{eq: est DkGk} for 

\quad \quad \quad \quad $t'\in [t,\cdots,T]$ with $\widehat{h}_k^{(t)}={h}_k^{(t)}$ }

\quad \quad \quad  {Apply Theorem~\ref{thm: optimal solution non-orth} over the time interval $[t,\cdots,T]$ with 

\quad\quad \quad \quad $\{{h}_k^{(t')}\leftarrow\widehat{h}_k^{(t')},\ {G}_k^{(t')}\leftarrow\widehat{G}_k^{(t')},\ \gamma^{(t)}\leftarrow \widehat{\gamma} \}$  and 

\quad\quad \quad \quad  residual DP constraint $\mathcal{R}_{\sf dp}-\mathcal{L}_g^{(t-1)}$}

\quad \quad  \quad  Broadcast optimized scaling factor $c^{(t)}$ to devices

\quad  \quad  \quad  {Update privacy loss as $\mathcal{L}_g^{(t)}=\mathcal{L}_g^{(t-1)}+\frac{(\sqrt{2}c^{(t)}\gamma^{(t)})^2}{N_0}$ }

\quad \quad {\bf end}

\quad \quad {\bf For each device:} $k=1,\cdots, K$

\quad\quad \quad Update local model by \eqref{eq: clipped local update}

\quad \quad  \quad  Receive $c^{(t)}$ and apply it to transmit \eqref{eq: clipped transmission NOMA}.

  \quad \quad {\bf end}

 {\bf end}
\caption{Online Scheme for NOMA}
\label{algorithm: online NOMA}
\end{algorithm}

\section{Numerical Results}\label{sec:simulation}

In this section, we evaluate the performance of the proposed schemes in order to gain insights into the impact of the DP constraints and into the benefits of adaptive power allocation.  We first consider a randomly generated synthetic dataset with $D_{\sf tot}=10000$ pairs $(\bu,v)$, where the covariates $\bu\in \mathbb{R}^{10}$ are drawn i.i.d. as $\mathcal{N}(0, \bI)$ and the label $v$ for each vector $\bu$ is obtained as $v=u(2)+3u(5)+0.2z_o$, where $u(d)$ is the $d$-th entry in vector $\bu$ and the observation noises $z_o\sim\mathcal{N}(0, 1)$ are i.i.d. across the samples \cite{seif2020wireless}.  Unless stated otherwise, the training samples are evenly distributed across the $K=10$ devices, so that the size of local data set is $D_k=1000$ for all $k$.   We consider ridge regression with the sample-wise loss function 
$f(\bw;\bu,v)=0.5\|\bw^{\sf T} \bu-v\|^2$ and the regularization function $R(\bw)=\|\bw\|^2$ with $\lambda=5\times10^{-5}$.
The PL parameter $\mu$ and smoothness parameter $L$ are computed as the smallest and largest eigenvalues of the data Gramian matrix $\bU^{\sf T}\bU/D_{\sf tot} +2\lambda\bI $, where $\bU=[\bu_1,\cdots,\bu_{D_{\sf tot}}]^{\sf T}$ is data matrix
of the data set.  The initial value for $\bw$ is set as an all-zero vector. We note that the (unique) optimal solution to the joint learning problem \eqref{eq: LP} is $\bw^*=(\bU^{\sf T}\bU +2D_{\sf tot}\lambda \bI )^{-1} \bU^{\sf T} \bv $, where $\bv=[v_1,\cdots, v_{D_{\sf tot}}]^{\sf T}$  is label vector.  We will also consider experiments with the MNIST data set at the end of this section.

Unless stated otherwise, the maximum SNR defined in \eqref{eq: SNR} is set to $\SNR_{\sf max}=30$ dB, and we consider the availability of $30$ communication blocks. Note that this implies $T=30/K=3$ iterations per device for OMA and $T=30$ iterations for NOMA.  Furthermore, the default DP settings are $\epsilon=20$ and $\delta=0.01$. 

As a benchmark, we consider a scheme that divides up the DP constraint equally across all iterations, i.e., it requires $({\sqrt{2}h_k^{(t)}\alpha_k^{(t)}\gamma^{(t)}}/{m_k^{(t)}})^2<\mathcal{R}_{\sf dp}/T$ for all $t=1,\cdots, T$ in lieu of constraint  \eqref{eq: OMA Local OPTb} and similarly for the constraint \eqref{NOMA OPTb}.  This yields 
\begin{align}
&{(\text{Static PA in OMA})} \nn\\
&\quad \alpha_k^{(t)}=\min\l\{\sqrt{\frac{N_o\mathcal{R}_{\sf dp}(\epsilon,\delta)}{2T(h_k^{(t)}\gamma^{(t)})^2}}, \frac{\sqrt{P}}{D_k G_k^{(t)}}\r\} ,  \\
&{(\text{Static PA in NOMA})}  \nn\\
&\quad c^{(t)}=\min\l\{\sqrt{\frac{N_o\mathcal{R}_{\sf dp}(\epsilon,\delta)}{2T(\gamma^{(t)})^2}},\sqrt{P}\min_k \frac{h_k^{(t)}}{D_kG_k^{(t)}}\r\}.
\end{align}
Another benchmark is set by the scheme that doe not impose the DP constraint \eqref{eq: OMA Local OPTb} and \eqref{NOMA OPTb}.  We adopt the normalized optimality gap $[F(\bw^{T+1})-F(\bw^*)]/F(\bw^*)$ as performance metric, and the offline results are averaged over 1000 channel realizations while online results are averaged over 100 channel realizations.

 \vspace{-8pt}
\subsection{Offline Optimization} 
We now focus on offline optimization by applying the optimal adaptive PA strategies in Theorems~\ref{thm: optimal solution TDMA} and \ref{thm: optimal solution non-orth}.  
For the channel model in \eqref{eq:channel model}, we set $\kappa=10$, and the channel correlation parameter is set as $\rho=1$, since this parameter has no discernible effect on the performance of offline strategies. 
We use the simple upper bounds $\gamma^{(t)}=2W \max_{(\bu,v)\in\mathcal{D}}L(\bu,v)$ and $G_k^{(t)}=2WL_k$, where $W\geq\|\bw\|$ is a bound on the norm $\|\bw\|$ (which can be in practice ensured via convex projection and is set to $W=3.2$ in our results); and $L(\bu,v)$ and $L_k$ are the Lipschitz smoothness constants of functions $f(\bw; \bu_n, v_n)$ and $F_k(\bw)$, respectively.

\begin{figure}[t]
\centering
\includegraphics[width=7.5cm]{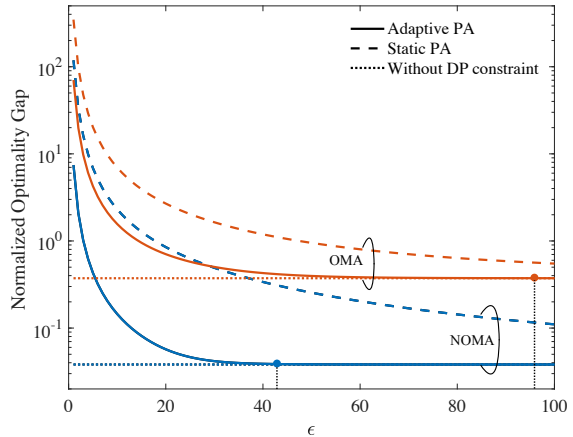}
\caption{Optimality gap versus DP privacy level $\epsilon$ (for $\delta=0.01$) for different power allocation (PA) schemes and for the scheme without DP constraint ($\SNR_{\sf max}=30$ dB).  }
\label{Fig: Epsilon}
\end{figure}

In Fig.~\ref{Fig: Epsilon}, we plot the normalized optimality gap as a function of the privacy level $\epsilon$. In the considered range of $\epsilon$, NOMA with either adaptive or static power allocation (PA) is seen to achieve better performance than OMA. Furthermore, adaptive PA achieves a significant performance gain over static PA under stringent DP constraints,  while the performance advantage of adaptive PA decreases as the DP constraint is relaxed, i.e., for larger values of $\epsilon$.   The figure also shows the threshold values of $\epsilon$ beyond which the privacy ``for free" conditions 
\eqref{eq: free dp condition} and \eqref{eq: condition free DP NOMA} are satisfied.  


\begin{figure}[t]
\centering
\includegraphics[width=7.5cm]{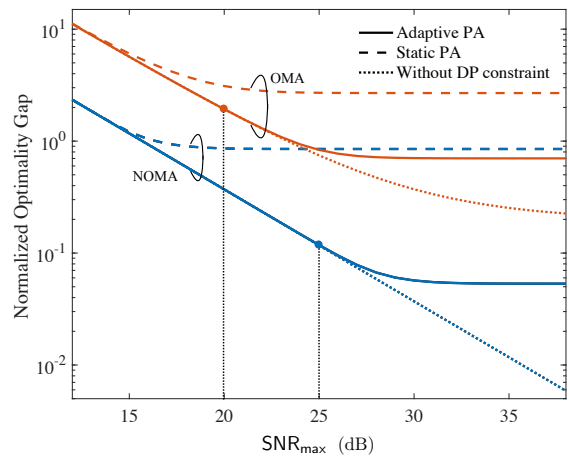}
\caption{Optimality gap versus $\SNR_{\sf max}$ for different power allocation (PA) schemes and for the scheme without DP constraint ($\epsilon=20$, $\delta=0.01$). }
\label{Fig: SNR}
\end{figure}

We now study the impact of the $\SNR_{\sf max}$ \eqref{eq: SNR} in Fig.~\ref{Fig: SNR}.  
The normalized optimality gap of all  schemes is seen to decrease with the SNR until the DP requirement becomes the performance bottleneck.   
While NOMA is confirmed to be generally advantageous over OMA, OMA with optimal PA can perform better than NOMA with static PA,  which emphasizes the importance of PA optimization, particularly in the high-SNR regime.  In a manner analogous to Fig.~\ref{Fig: Epsilon}, the plot also marks the maximum SNR levels for which the privacy ``for free" condition \eqref{eq: free dp condition} and \eqref{eq: condition free DP NOMA} are satisfied. 
%

\begin{figure}[t]
\centering
\includegraphics[width=7.5cm]{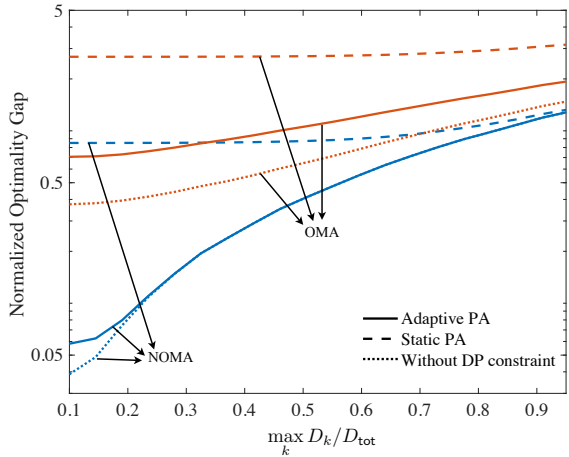}
\caption{Optimality gap versus data set heterogeneity parameter $\max_k D_k/D_{\sf tot}$ for different power allocation (PA) schemes and the scheme without DP constraint ($\epsilon=20$, $\delta=0.01$, $\SNR_{\sf max}=30$ dB).}
\label{Fig: Dk}
\end{figure}

Fig.~\ref{Fig: Dk} plots the normalized optimality gap versus a measure of the heterogeneity of the data sets.  To this end, one of the devices is allocated a lager value $D_k$, while the remaining data points are evenly distributed to the other devices.  The ratio $\max_k D_k/D_{\sf tot}$ varies from $0.1$ (uniformly distributed) to $0.95$ (highly skewed).  Increasing data set heterogeneity generally affects negatively all schemes,  even in the absence of privacy constraints.  Nevertheless, the heterogeneity of $D_k$ has a stronger impact for NOMA  than for OMA due to gradient alignment condition \eqref{eq: gradient alignment}.  In particular,  for NOMA,  the power constraint becomes the performance bottleneck as the ratio $\max D_k/D_{\sf tot}$ increases, and the performance of adaptive PA converges first to that without the DP constraint and then to that of static PA.

  \vspace{-8pt}
 \subsection{Online Optimization}
 
 We now turn to the heuristic online optimization methods proposed in Algorithms 1 and 2.  For the channel model in \eqref{eq:channel model}, we set $\kappa=5$ and $\rho=0$. Note that channel prediction is possible due to the non-zero Rician factor. The maximum value of $\|\bw\|$ is set as $W=10$, which is ensured by convex projection.  Unless stated otherwise, we set the clipping threshold as $\widehat{\gamma}=20$.

\begin{figure}[t]
\centering
\includegraphics[width=7.5cm]{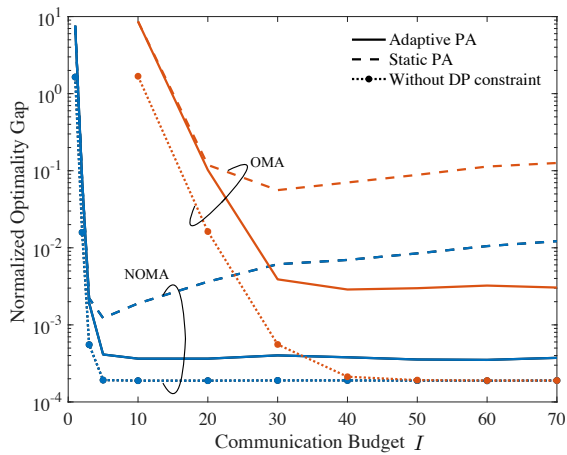}
\caption{Optimality gap versus communication budget $I$ for different power allocation (PA) schemes and the scheme without DP constraint ($\epsilon=20$, $\delta=0.01$, $\SNR_{\sf max}=30$ dB, $\kappa=5$, $\rho=0$).}
\label{Fig: commbudget}
\end{figure}

In Fig.~\ref{Fig: commbudget}, we study the impact of the communication budget in terms of number of communication blocks $I$.  With conventional static PA, there exists an optimal communication budget under privacy constraints.  This is because more communication blocks may cause an increase in privacy loss (see also \cite{wei2020federated}). 
In contrast, increasing the communication budget always benefits adaptive PA, which is able to properly allocate power across the communication blocks.  
Furthermore, without DP constraint, the performance of OMA converges to that of NOMA when the communication budget $I$ is large; while, under privacy constraints, NOMA  retains performance advantages even with a large $I$.  

\begin{figure}[t]
\centering
\includegraphics[width=7.5cm]{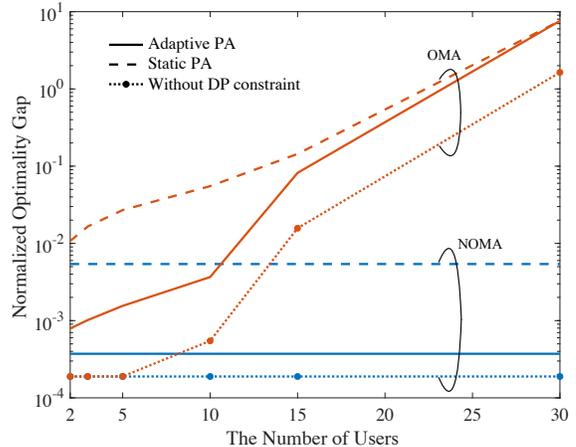}
\caption{Optimality gap versus the number of user $K$ for different power allocation (PA) schemes and for the scheme without DP constraint ($\epsilon=20$, $\delta=0.01$, $\SNR_{\sf max}=30$ dB, $\kappa=5$, $\rho=0$).}
\label{Fig: numuser}
\end{figure}

Fig.~\ref{Fig: numuser} plots the normalized optimality gap versus the number of users. It shows that increasing the number of users has a negative effect on OMA, but it causes no harm to NOMA.  This emphasizes the spectral efficiency of NOMA in wireless edge learning.  Furthermore, under OMA, a larger number of users implies fewer iterations, and thus less information leakage of each user, decreasing the performances gain of adaptive PA.  Specifically, when $K=30$, a simple iteration $T=1$ is carried out by OMA, and adaptive PA is equivalent to static PA.   

\begin{figure}[t]
\centering
\includegraphics[width=7.5cm]{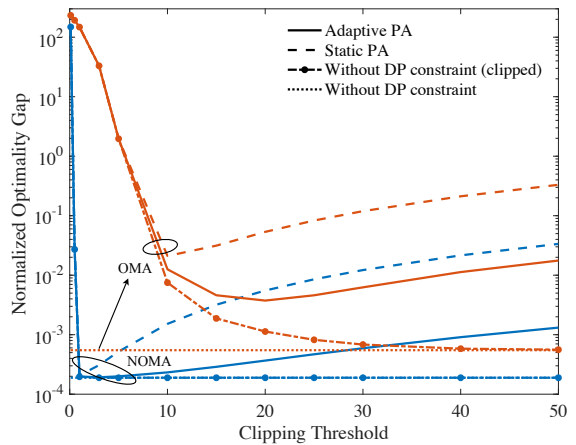}
\caption{Optimality gap versus the clipping threshold $\widehat{\gamma}$ for different power allocation (PA) schemes and for the schemes without DP constraint ($\epsilon=20$, $\delta=0.01$, $\SNR_{\sf max}=30$ dB, $T=30$).}
\label{Fig: ClipValue}
\end{figure}

We now study the impact of the clipping threshold $\widehat{\gamma}$ for the gradient in Fig.~\ref{Fig: ClipValue}.  To show the impact of clipping, we also plot the performance with clipped local updates without the DP constraint for both OMA and NOMA.  Without DP constraint, the larger clipping threshold incurs a smaller distortion of the gradients, which benefits the learning performance.  However, under the DP constraint, increasing the clipping threshold beyond a given value degrades the performance, since ensuring privacy requires a more pronounced scaling down of the transmitted signals.  This indicates the importance of selecting a threshold $\widehat{\gamma}$ that strikes a balance between learning performance and privacy. 

{\color{black}
 \subsection{MNIST Data Set}
We now consider the problem of classification on the MNIST data set via  multinomial logistic regression with quadratic regularization. Accordingly, the global loss function is given as the regularized cross-entropy loss
\begin{align*}
F(\bw )=\frac{1}{D_{\sf tot}}
\sum_{(\bu,v)\in \cD} \sum_{c=1}^C {\bf 1}\{v=c\}&\log \frac{\exp(\bw_c ^{\sf T} \bu )}{ \sum_{j=1}^C\exp(\bw_j ^{\sf T} \bu )}  \\
& \qquad +\lambda \sum_{c=1}^C\|\bw_c\|^2,
\end{align*}
where $C=10$ represents the total number of classes of handwritten digits; $\bu$ is data image extended to include a bias term; and the model parameter $\bw$, with dimension $7650$,  is comprised of the per-class vectors $\{\bw_c\}_{c=1}^C$.  We set $\lambda=0.01$, the maximum value of $\|\bw\|$ to $W=10$, the clipping threshold as $\widehat{\gamma}=40$, and $\SNR_{\sf max}=13$ dB. The smoothness parameter $L$ and strongly convex parameter $\mu$ are treated as hyper-parameter and selected via validation as  $\mu=0.3$ and $L=2.5$.
For $\epsilon=5$ and $\delta=0.01$, 
Fig. 9 plots the training cross-entropy loss and the probability of error on the test set versus the value of communication budget $I$ for OMA. Adaptive PA is seen to significantly outperform static PA both in terms of training loss and test error. Similar results can be obtained for NOMA.  
\begin{figure}[t]
\centering
\subfigure[Training cross-entropy loss versus communication budget $I$.]{
\label{Fig: TrainLoss}
\includegraphics[width=7.5cm]{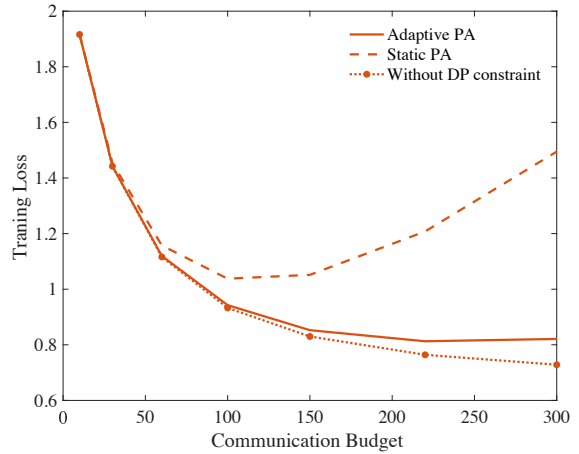}}
\subfigure[Test probability of error versus communication budget $I$.]{
\label{Fig: TestError}
\includegraphics[width=7.5cm]{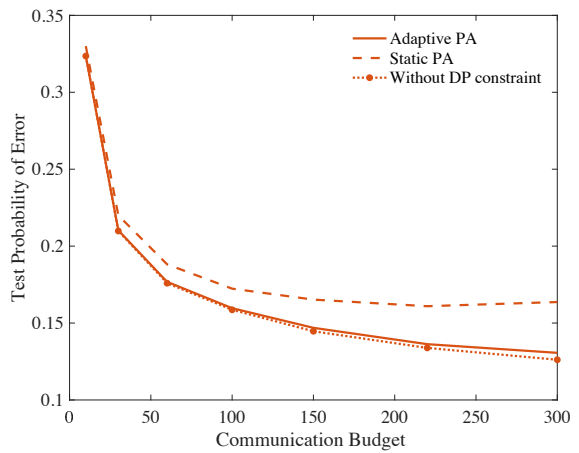}}
\caption{Training loss and test error for different power allocation (PA) schemes on the MNIST dat set for OMA ($\epsilon=5$, $\delta=0.01$, $\SNR_{\sf max}=13$ dB, $\kappa=5$, $\rho=0$).}
\vspace{-6mm}
\end{figure}
}

\section{Conclusions}\label{sec: conclusions}
In this paper, we have considered differentially private wireless federated learning via the direct, uncoded, transmission of gradient from devices to edge server.  The proposed approach is based on adaptive PA schemes that are optimized to minimize the learning optimality gap under privacy and power constraints. First, offline optimization problems are separately formulated for OMA and NOMA, which are converted to  convex programs.  The optimal PA, obtained in closed form, adapts the power along the iterations, outperforming static PA assumed in prior works.  Furthermore, a heuristic online approach is proposed that leverages iterative one-step-ahead optimization based on the offline result and predicted CSI.  

The analysis in this paper proved that privacy can be obtained ``for free", that is without affecting the learning performance, as long as the privacy constraint level is below a threshold that decreases with the SNR.  Our analytical results also demonstrate that it is generally suboptimal to devote part of the transmitted power to actively add noise to the local updates. This is unlike the standard scenario with ideal communication, in which adding noise is essential to ensure DP constraints.  Via numerical results, we have finally shown that techniques that leverage over-the-air computing provide significant benefits over conventional OMA protocols under DP constraints.  This is not a prior evident, since, with NOMA, devices transmit more frequently, and hence may leak more information. 

{\color{black}
We note that the power control policy based on channel inversion for all the devices was proven to be suboptimal in the scenario of over-the-air computing without DP constraint.  In fact, channel inversion can incur noise amplification by adapting the power to the device with worst channel condition \cite{zhu2019broadband,cao2020optimized}. However, this may not be the case under DP constraints since noise amplification benefits privacy. As a possible extension of the current work, it would be  interesting to study the optimization of the threshold for channel inversion so as to maximize the learning performance under privacy and power constraints.}
As directions for future work, the threat model could also include  ``honest-but-curious" edge devices, which would generally incur larger DP loss.  The study could be further generalized to other network topologies including multi-hop device-to-device (D2D) networks.  Another interesting direction is to consider the implementation of digital transmission where quantization introduces additional privacy preserving mechanism on top of the channel noise.  {\color{black}
It would also be interesting to investigate the effect of clipping in terms of convergence as in \cite{chen2020understanding,zhang2019gradient}, and to address the convergence properties of the proposed online scheme.
}



\appendix

{\color{black}
\subsection{Proof of Lemma~\ref{lemma: privacy constraint}}\label{proof: DP constraint}
To start, we denote  as $\by_k=[\by_k^{(1)},\cdots,\by_k^{(T)}]$  the $T$ successive received signals from device $k$, and $m_k^{(t)}=\sqrt{(h_k^{(t)}\alpha_k^{(t)}\sigma_k^{(t)})^2+N_0}$ is the standard deviation of the effective noise in $\by_k^{(t)}$.   According to the definition of DP loss given in \eqref{def: privacy loss}, for the $k$-th device, the privacy loss after $T$ iterations can be represented as
\begin{align}
&\mathcal{L}_{\mathcal{D},\mathcal{D}'}(\by_k)=\ln\l(\prod_{t=1}^T \frac{P\Big[ \by_k^{(t)}\big| \by_k^{(t-1)},\cdots, \by_k^{(1)},\mathcal{D}_k\Big]}{P\Big[ \by_k^{(t)}\big| \by_k^{(t-1)},\cdots, \by_k^{(1)},\mathcal{D}'_k\Big]}\r)\nn\\
&=\sum_{t=1}^T\ln\l( \frac{P\Big[ \by_k^{(t)}\big| \by_k^{(t-1)},\cdots, \by_k^{(1)},\mathcal{D}_k\Big]}{P\Big[ \by_k^{(t)}\big| \by_k^{(t-1)},\cdots, \by_k^{(1)},\mathcal{D}'_k\Big]}\r)\nn\\
&=\sum_{t=1}^T\ln\l(\frac{\exp\l(-\frac{ \|\by_k^{(t)}-h_k^{(t)}\alpha_k^{(t)}D_k\nabla F_k(\bw^{(t)};\mathcal{D}_k)\|^2}{2(m_k^{(t)})^2}\r)}{\exp\l(-\frac{ \|\by_k^{(t)}-h_k^{(t)}\alpha_k^{(t)}D_k\nabla F_k(\bw^{(t)};\mathcal{D}'_k)\|^2}{2(m_k^{(t)})^2}\r)}\r)\nn\\
&=\sum_{t=1}^T\ln\l(\frac{\exp\l(-\frac{\| {\br}_k^{(t)}\|^2}{2(m_k^{(t)})^2}\r)}{\exp\l(-\frac{\|{\bf r}_k^{(t)}+\bv_k^{(t)}\|^2}{2(m_k^{(t)})^2}\r)}\r), \nn
\end{align}
where ${\bf r}_k^{(t)}\sim\mathcal{N}(0,{(m_k^{(t)})^2}\bI)$ represents the effective noise, and we set
\begin{equation*}
\bv_k^{(t)}= {h_k^{(t)}\alpha_k^{(t)}} \Big[\!\!\sum_{(\bu,v)\in\mathcal{D}_k}\!\!\!\!\!\! \nabla f\l(\bw^{(t)};\bu,v\r)-\!\!\!\!\!\sum_{(\bu,v)\in\mathcal{D}_k'}\!\!\!\!\!\! \nabla f\l(\bw^{(t)};\bu,v\r)\!\!\Big]
\end{equation*}
 with $\|\bv_k^{(t)}\|=\Delta_k^{(t)}$.  Following \cite[Appendix A]{dwork2014algorithmic}, we can then bound privacy violation probability  
\begin{align}
&\Pr\l(\l|\sum_{t=1}^T\frac{2({\bf r}_k^{(t)})^{\sf T}\bv_k^{(t)}+\|\bv_k^{(t)}\|^2}{2(m_k^{(t)})^2}\r|>\epsilon\r)\nn\\
&\overset{(a)}{\leq}\Pr\l(\l|\sum_{t=1}^T\frac{({\bf r}_k^{(t)})^{\sf T}\bv_k^{(t)}}{(m_k^{(t)})^2}\r|>\epsilon-\sum_{t=1}^T\frac{\|\bv_k^{(t)}\|^2}{2(m_k^{(t)})^2}\r)\nn\\
&=2\Pr\l(\sum_{t=1}^T\frac{({\bf r}_k^{(t)})^{\sf T}\bv_k^{(t)}}{(m_k^{(t)})^2}>\epsilon-\sum_{t=1}^T\frac{\|\bv_k^{(t)}\|^2}{2(m_k^{(t)})^2}\r)\nn\\
&\overset{(b)}{\leq} 2\frac{\sqrt{\sum_{t=1}^T\Big(\frac{\Delta_k^{(t)}}{m_k^{(t)}}\Big)^2}}{\sqrt{2\pi}\l[\epsilon-\sum_{t=1}^T\frac{1}{2}\Big(\frac{\Delta_k^{(t)}}{m_k^{(t)}}\Big)^2\r]}\nn\\
& \qquad \qquad\times \exp\l(-\frac{\l[\epsilon-\sum_{t=1}^T\frac{1}{2}\Big(\frac{\Delta_k^{(t)}}{m_k^{(t)}}\Big)^2\r]^2}{2\sum_{t=1}^T\Big(\frac{\Delta_k^{(t)}}{m_k^{(t)}}\Big)^2}\r), \label{eq: DP loss ub}
\end{align}
where $(a)$ is obtained by using $\Pr(X<-\epsilon-b)\leq\Pr(X<-\epsilon+b)$ for an arbitrary $b\geq 0$,  and $(b)$ comes from the following bound on the tail probability of Gaussian distribution $X\sim\mathcal{N}\l(0,\sigma^2\r)$: $\Pr(X>s)=\frac{1}{\sigma\sqrt{2\pi}}\int_s^{\infty}\exp(-\frac{x^2}{2\sigma^2})dx\leq\frac{1}{\sigma\sqrt{2\pi}}\int_s^{\infty}\frac{x}{s}\exp(-\frac{x^2}{2\sigma^2})dx=\frac{\sigma}{s\sqrt{2\pi}}\exp\l(-\frac{s^2}{2\sigma^2}\r)$.

Letting $q=\frac{\epsilon-\sum_{t=1}^T\frac{1}{2}(\Delta_k^{(t)}/m_k^{(t)})^2}{\sqrt{2\sum_{t=1}^T(\Delta_k^{(t)}/m_k^{(t)})^2}}$ and using \eqref{eq: DP loss ub}, the DP condition is implied by the inequality 
\begin{align}\label{eq: DP last bound}
\Pr(|\mathcal{L}_{\mathcal{D},\mathcal{D}'}(\by_k)|>\epsilon)\leq \frac{1}{q\sqrt{\pi}}e^{-q^2}<\delta.
\end{align}
Finally, defining the function $\mathcal{C}(x)=\sqrt{\pi}{x}e^{x^2}$ and utilizing its monotonicity yields the desired result. 

}

\subsection{Proof of Lemma~\ref{lemma: convergence}}\label{proof: convergence}

Under Assumption~\ref{assumption: Lipschitz}, we have the following equality
\begin{align}
 &F\big(\bw^{(t)}\big)\leq F\big(\bw^{(t-1)}\big)+ \l[\nabla F\big(\bw^{(t-1)}\big)\r]^{\sf T}\l[\bw^{(t)}-\bw^{(t-1)}\r] \nn\\
 &\qquad \qquad\qquad\qquad \qquad\qquad \qquad +\frac{L}{2}\l\|\bw^{(t)}-\bw^{(t-1)}\r\|^2 \nn\\
 &\qquad\qquad =F\big(\bw^{(t-1)}\big) \!-\!\eta\l[\nabla F\big(\bw^{(t-1)}\big)\r]^{\sf T} \!\! \times \bigg[\nabla F\big(\bw^{(t-1)}\big) \nn\\
 &\qquad\qquad \quad +\frac{1}{D_{\sf tot}}\sum_{k=1}^K\Big[{\bf n}_k^{\l(t-1\r)} + \(h_k^{(t-1)}\alpha_k^{(t-1)}\r)^{-1}\bz_k^{(t-1)}\Big]\bigg]\nn\\
&\qquad\qquad \quad +\frac{L\eta^2}{2}\bigg\|\nabla F\big(\bw^{(t-1)}\big)+\frac{1}{D_{\sf tot}}\sum_{k=1}^K \Big[{\bf n}_k^{\l(t-1\r)} + \nn\\
&\qquad\qquad \quad \qquad\qquad \qquad \quad  \big(h_k^{(t-1)}\alpha_k^{(t-1)}\big)^{-1}\bz_k^{(t-1)}\Big]\bigg\|^2.\nn
\end{align}
By taking the expectation over the additive noise on both sides of the above inequality, we obtain
\begin{align}
 \E\l[F\big(\bw^{(t)}\big)\r]  \leq& F\big(\bw^{(t-1)}\big)-\eta\l[\l(1-\frac{L\eta}{2}\r)\l\| \nabla F\big(\bw^{(t-1)}\big)\r\|^2\r]\nn\\
 & \quad \quad +\frac{L\eta^2d}{2D_{\sf tot}^2}\sum_{k=1}^K(\sigma_k^{(t-1)})^2+\bigg(\frac{\sqrt{N_0}}{h_k^{(t-1)}\alpha_k^{(t-1)}}\bigg)^2 \nn\\
=&F\big(\bw^{(t-1)}\big)-\frac{1}{2L}\l\| \nabla F\big(\bw^{(t-1)}\big)\r\|^2 \nn\\
& \qquad\qquad+\frac{d}{2LD_{\sf tot}^2}\sum_{k=1}^K \bigg(\frac{m_k^{(t-1)}}{h_k^{(t-1)}\alpha_k^{(t-1)}}\bigg)^2, \nn
\end{align}
where the equality follows by Lemma~\ref{lemma: privacy constraint} and by setting $\eta={1}/{L}$. 

Subtracting the optimal value $F^*$ at both sides yields
\begin{align}
 &\E\l[F\big(\bw^{(t)}\big)\r]-F^*  \nn\\
 & \quad\leq F\big(\bw^{(t-1)}\big)-F^*-\frac{1}{2L}\l\| \nabla F\big(\bw^{(t-1)}\big)\r\|^2 \nn\\
 &\qquad \qquad \qquad +\frac{d}{2LD_{\sf tot}^2}\sum_{k=1}^K \l(\frac{m_k^{(t-1)}}{h_k^{(t-1)}\alpha_k^{(t-1)}}\r)^2 \nn \\
 &\quad\leq \l(1-\frac{\mu}{L}\r)\l(F\big(\bw^{(t-1)}\big)-F^*\r)+\frac{d}{2LD_{\sf tot}^2} \nn\\
 & \qquad \qquad \qquad\qquad \qquad\times\sum_{k=1}^K \l(\frac{m_k^{(t-1)}}{h_k^{(t-1)}\alpha_k^{(t-1)}}\r)^2,
\end{align}
 where the last step follows from Assumption~\ref{assumption: PL}.
Then, the desired result yields by applying above inequality repeatedly through $T$ iterations and taking expectation over all the additive~noises. 


\subsection{Proof of Theorem~\ref{thm: optimal solution TDMA}}\label{proof: optimal solution TDMA}
We start by making the change of variables
\begin{align} \vspace{-10pt}
  a_k^{(t)}=(\sigma_k^{(t)})^2+ {N_0}/ (h_k^{(t)}\alpha_k^{(t)})^{2}, \quad b_k^{(t)}=({\alpha_k^{(t)}})^{-2},   \vspace{-10pt}
\end{align} 
so that  the original variables can be written as
\begin{align} 
(\sigma_k^{(t)})^2&=  a_k^{(t)} - ({\sqrt{N_0}}/{h_k^{(t)}})^2 b_k^{(t)} \geq 0, \ (\alpha_k^{(t)})^2={1}/{b_k^{(t)}}>0. \label{eq: positive power}
\end{align}
By including the constraints \eqref{eq: positive power}, we now obtain the equivalent local problem {({\bf OMA Local Opt. 2})} 
\begin{align*} 
\quad \min_{\{{a_k^{(t)},b_k^{(t)} \}}}&\quad 
\sum_{t=1}^{T} \l(1-\frac{\mu}{L}\r)^{-t}  a_k^{(t)} \\
{\rm s.t.} \quad &\sum_{t=1}^{T} {(\sqrt{2}\gamma^{(t)})^2}/{  a_k^{(t)}}\leq \mathcal{R}_{\sf dp},\\
&(D_kG_k^{(t)})^2+d a_k^{(t)} \!- \! \l[d ({\sqrt{N_0}}/{h_k^{(t)}})^2  +P\r]b_k^{(t)} \leq 0, \nn\\
 &\qquad  \qquad \qquad \qquad \qquad  \qquad \qquad \qquad \qquad\qquad \forall t, \\
&  a_k^{(t)} - ({\sqrt{N_0}}/{h_k^{(t)}})^2  b_k^{(t)} \geq 0, \quad \forall t,  \\
&  a_k^{(t)} \geq 0, \quad
  b_k^{(t)} \geq0,\qquad \forall t, 
 \end{align*}
which is a convex optimization problem.  To solve it, the partial Lagrange function is defined as 
 \begin{align}
\mathcal{L}=\sum_{t=1}^{T} &\l(1-\frac{\mu}{L}\r)^{-t}a_k^{(t)}+\zeta\l( \sum_{t=1}^{T}\frac{(\sqrt{2}\gamma^{(t)})^2}{a_k^{(t)}} -\mathcal{R}_{\sf dp}\r) \nn\\
& +\sum_{t=1}^{T}  \xi^{(t)}\l(\l({\sqrt{N_0}}/{h_k^{(t)}}\r)^2 b_k^{(t)} -a_k^{(t)}\r)\nn\\
&+ \sum_{t=1}^{T} \beta^{(t)} \bigg( (D_kG_k^{(t)})^2+d a_k^{(t)}  \nn\\
& \qquad \qquad \qquad-  \Big[d \big({\sqrt{N_0}}/{h_k^{(t)}}\big)^2 +P\Big]b_k^{(t)}\bigg)  ,
 \end{align}
where $\zeta\geq0$, $\beta^{(t)}\geq0$,  and $\xi^{(t)}\geq0$ are the Lagrange multipliers associated respectively with the DP constraint, transmit power constraints and non-negative parameter constraints.    Then applying the KKT conditions leads to the following necessary and sufficient conditions
  \begin{subequations}\label{eq: KKT OMA}
 \begin{align}
&\frac{\partial \mathcal{L}}{\partial( a_k^{(t)})_{\sf opt}}=\l(1-\frac{\mu}{L}\r)^{-t}-\zeta_{\sf opt}{(\sqrt{2}\gamma^{(t)})^2}\l((a_k^{(t)})_{\sf opt}\r)^{-2} \nn\\
& \qquad \qquad \quad \qquad   \qquad  +(\beta^{(t)})_{\sf opt}d-(\xi^{(t)})_{\sf opt} =0, \label{eq: KKT lanrange xt} \\
&\frac{\partial \mathcal{L}}{\partial (b_k^{(t)})_{\sf opt}}=-(\beta^{(t)})_{\sf opt}\l[d ({\sqrt{N_0}}/{h_k^{(t)}})^2 +P\r] \nn\\
& \qquad \qquad \quad \qquad   \qquad+(\xi^{(t)})_{\sf opt}({\sqrt{N_0}}/{h_k^{(t)}})^2=0,\label{eq: KKT lanrange yt}\\
&\zeta_{\sf opt}\l( \sum_{t=1}^{T}\frac{(\sqrt{2}\gamma^{(t)})^2}{(a_k^{(t)})_{\sf opt}} -\mathcal{R}_{\sf dp}\r)=0,\label{eq: KKT DP}\\
& (\beta^{(t)})_{\sf opt} \bigg( (D_kG_k^{(t)})^2+d (a_k^{(t)})_{\sf opt}-  \Big[d \big({\sqrt{N_0}}/{h_k^{(t)}}\big)^2 +P\Big]\nn\\
& \qquad \qquad \quad \qquad  \qquad \qquad  \quad \qquad\times(b_k^{(t)})_{\sf opt}\bigg)=0, \label{eq: KKT power constraint}\\
& (\xi^{(t)})_{\sf opt}\Big[\big({\sqrt{N_0}}/{h_k^{(t)}}\big)^2 (b_k^{(t)})_{\sf opt} -(a_k^{(t)})_{\sf opt}\Big]=0, \label{eq: KKT non zero} \\
&  \sum_{t=1}^{T}\frac{(\sqrt{2}\gamma^{(t)})^2}{(a_k^{(t)})_{\sf opt}} -\mathcal{R}_{\sf dp}\leq 0, \label{eq: KKT ineq DP}\\
& (D_kG_k^{(t)})^2+d (a_k^{(t)})_{\sf opt} \!-\!  \Big[d \big({\sqrt{N_0}}/{h_k^{(t)}}\big)^2 +P\Big](b_k^{(t)})_{\sf opt} \nn\\
& \qquad \qquad \qquad \qquad \qquad \qquad \qquad  \qquad  \qquad \leq0,  \label{eq: KKT ineq power} \\
&({\sqrt{N_0}}/{h_k^{(t)}}\big)^2 (b_k^{(t)})_{\sf opt} -(a_k^{(t)})_{\sf opt} \leq 0.  \label{eq: KKT ineq non zero}
 \end{align} 
 \end{subequations}
According to \eqref{eq: KKT lanrange yt}, we have the following equality for the optimal solutions $(\beta^{(t)})_{\sf opt}$ and $(\xi^{(t)})_{\sf opt}$
\begin{align}
(\xi^{(t)})_{\sf opt}=\frac{d ({\sqrt{N_0}}/{h_k^{(t)}})^2 +P}{({\sqrt{N_0}}/{h_k^{(t)}})^2}(\beta^{(t)})_{\sf opt}. \label{eq: equality xi and beta}
\end{align}
Plugging the above result into \eqref{eq: KKT lanrange xt} and \eqref{eq: KKT non zero}, respectively, we obtain
\begin{align}
&\l(1-\frac{\mu}{L}\r)^{-t}-\zeta_{\sf opt}{(\sqrt{2}\gamma^{(t)})^2}\l((a_k^{(t)})_{\sf opt}\r)^{-2}\nn\\
&\qquad \qquad \qquad\qquad\qquad\qquad-(\beta^{(t)})_{\sf opt} \frac{P (h_k^{(t)})^2 }{(\sqrt{N_0})^2}=0 \label{eq: only beta KKT xt}\\
&(\beta^{(t)})_{\sf opt} \frac{d ({\sqrt{N_0}}/{h_k^{(t)}})^2 +P}{({\sqrt{N_0}}/{h_k^{(t)}})^2} \bigg(\l({\sqrt{N_0}}/{h_k^{(t)}}\r)^2 (b_k^{(t)})_{\sf opt} \nn\\
&\qquad\qquad\qquad\qquad\qquad\qquad\qquad-(a_k^{(t)})_{\sf opt}\bigg)=0. \label{eq: only beta non zero}
\end{align}
Combining \eqref{eq: only beta non zero} and \eqref{eq: KKT power constraint}, we get the following equation
\begin{align}
(\beta^{(t)})_{\sf opt}\l((D_kG_k^{(t)})^2 - P( {h_k^{(t)}}/{\sqrt{N_0}})^2 (a_k^{(t)})_{\sf opt} \r)=0.  \label{eq: min value xt}
\end{align}
Constraints \eqref {eq: KKT ineq power} and \eqref{eq: KKT ineq non zero} define the minimum and maximum values of $(b_k^{(t)})_{\sf opt}$ in terms of $(a_k^{(t)})_{\sf opt}$.  Accordingly, the minimum value of $(b_k^{(t)})_{\sf opt}$ should be no larger than that of the maximum value, which yields the following lower bound on $(a_k^{(t)})_{\sf opt}$: 
\begin{equation}
(a_k^{(t)})_{\sf opt}\geq  (D_kG_k^{(t)})^2 ({\sqrt{N_0}}/{h_k^{(t)}})^2 /P. \label{eq: solution by power const}
\end{equation}
 In this case, the power is fully utilized for transmitting the local gradient 

Furthermore, from \eqref{eq: min value xt}, we have the equality $(\beta^{(t)})_{\sf opt}=0$ if $(a_k^{(t)})_{\sf opt}>  (D_kG_k^{(t)})^2 ({\sqrt{N_0}}/{h_k^{(t)}})^2 /P$, thereby the other solution of $(a_k^{(t)})_{\sf opt}$ is obtained by solving \eqref{eq: only beta KKT xt} as 
\begin{equation}
(a_k^{(t)})_{\sf opt}=\sqrt{2\zeta_{\sf opt}}(1-\mu/L)^{t/2}\gamma^{(t)}. \label{eq: solution by dp}
\end{equation}
 Combing \eqref{eq: solution by power const} and \eqref{eq: solution by dp}, the solution of $(a_k^{(t)})_{\sf opt}$ is  
\begin{align}
&(a_k^{(t)})_{\sf opt}=\max\Big\{\sqrt{2\zeta_{\sf opt}}(1-\mu/L)^{t/2}\gamma^{(t)}, \nn\\
&\qquad \qquad \qquad \qquad \qquad (D_kG_k^{(t)})^2 ({\sqrt{N_0}}/{h_k^{(t)}})^2 /P\Big\}, 
\end{align}
and the value of $\zeta_{\sf opt}$ can be obtained by bisection search to satisfy the equality of \eqref{eq: KKT ineq DP}. Specifically, we have $\zeta_{\sf opt}=0$ if $\sum_{t=1}^T (\sqrt{2}\gamma^{(t)} h_k^{(t)})^2 P/(\sqrt{N_0}D_kG_k^{(t)})^2<\mathcal{R}_{\sf dp}$.  With the value of  $(a_k^{(t)})_{\sf opt}$, the solution of $(b_k^{(t)})_{\sf opt}$ can be obtained by using \eqref {eq: KKT ineq power} and \eqref{eq: KKT ineq non zero}, which are satisfied by arbitrary value within the range 
\begin{align}\label{eq: range of b}
\frac{(D_kG_k^{(t)})^2+d (a_k^{(t)})_{\sf opt}}{d ({\sqrt{N_0}}/{h_k^{(t)}})^2 +P}\leq (b_k^{(t)})_{\sf opt} \leq \frac{ (h_k^{(t)})^2(a_k^{(t)})_{\sf opt}}{N_0}.
\end{align}
Then, the desired result in the theorem is obtained by reverting to the original variables using  \eqref{eq: positive power}. Specifically,  the optimal solution to minimize the transmit power is attained by the maximum value of $(b_k^{(t)})_{\sf opt}$ in \eqref{eq: range of b}.

\bibliographystyle{ieeetr}

\begin{thebibliography}{10}

\bibitem{park2019wireless}
J.~Park, S.~Samarakoon, M.~Bennis, and M.~Debbah, ``Wireless network
  intelligence at the edge,'' {\em Proc. IEEE}, vol.~107, no.~11,
  pp.~2204--2239, 2019.

\bibitem{zhou2019edge}
Z.~Zhou, X.~Chen, E.~Li, L.~Zeng, K.~Luo, and J.~Zhang, ``Edge intelligence:
  Paving the last mile of artificial intelligence with edge computing,'' {\em
  Proc. IEEE}, vol.~107, no.~8, pp.~1738--1762, 2019.

\bibitem{zhu2020toward}
G.~Zhu, D.~Liu, Y.~Du, C.~You, J.~Zhang, and K.~Huang, ``Toward an intelligent
  edge: wireless communication meets machine learning,'' {\em IEEE Commun.
  Mag.}, vol.~58, no.~1, pp.~19--25, 2020.

\bibitem{li2020federated}
T.~Li, A.~K. Sahu, A.~Talwalkar, and V.~Smith, ``Federated learning:
  Challenges, methods, and future directions,'' {\em IEEE Signal Process.
  Mag.}, vol.~37, no.~3, pp.~50--60, 2020.

\bibitem{kairouz2019advances}
P.~Kairouz, H.~B. McMahan, B.~Avent, A.~Bellet, M.~Bennis, A.~N. Bhagoji,
  K.~Bonawitz, Z.~Charles, G.~Cormode, R.~Cummings, {\em et~al.}, ``Advances
  and open problems in federated learning,'' {\em [Online]. Available:
  https://arxiv.org/pdf/1912.04977.pdf}, 2019.

\bibitem{melis2019exploiting}
L.~Melis, C.~Song, E.~De~Cristofaro, and V.~Shmatikov, ``Exploiting unintended
  feature leakage in collaborative learning,'' in {\em IEEE Symp. Secur. and
  Privacy (SP)}, pp.~691--706, IEEE, 2019.

\bibitem{fredrikson2015model}
M.~Fredrikson, S.~Jha, and T.~Ristenpart, ``Model inversion attacks that
  exploit confidence information and basic countermeasures,'' in {\em Proc. ACM
  SIGSAC Conf. Comput. Commun. Secur.}, (Denver, USA), Oct. 2015.

\bibitem{bhowmick2018protection}
A.~Bhowmick, J.~Duchi, J.~Freudiger, G.~Kapoor, and R.~Rogers, ``Protection
  against reconstruction and its applications in private federated learning,''
  {\em [Online]. Available: https://arxiv.org/pdf/1812.00984.pdf}, 2018.

\bibitem{dwork2014algorithmic}
C.~Dwork, A.~Roth, {\em et~al.}, ``The algorithmic foundations of differential
  privacy,'' {\em Foundations and Trends{\textregistered} in Theoretical
  Computer Science}, vol.~9, no.~3--4, pp.~211--407, 2014.

\bibitem{nazer2007computation}
B.~Nazer and M.~Gastpar, ``Computation over multiple-access channels,'' {\em
  IEEE Trans. Inf. Theory}, vol.~53, no.~10, pp.~3498--3516, 2007.

\bibitem{zhu2019broadband}
G.~Zhu, Y.~Wang, and K.~Huang, ``Broadband analog aggregation for low-latency
  federated edge learning,'' {\em IEEE Trans. Wireless Commun.}, 2019.

\bibitem{sery2020analog}
T.~Sery and K.~Cohen, ``On analog gradient descent learning over multiple
  access fading channels,'' {\em IEEE Trans. Signal Process.}, vol.~68,
  pp.~2897--2911, 2020.

\bibitem{amiri2020machine}
M.~M. Amiri and D.~G{\"u}nd{\"u}z, ``Machine learning at the wireless edge:
  Distributed stochastic gradient descent over-the-air,'' {\em IEEE Trans.
  Signal Process.}, vol.~68, pp.~2155--2169, 2020.

\bibitem{zhang2020gradient}
N.~Zhang and M.~Tao, ``Gradient statistics aware power control for over-the-air
  federated learning,'' {\em [Online]. Available:
  https://arxiv.org/pdf/2003.02089.pdf}, 2020.

\bibitem{yang2020federated}
K.~Yang, T.~Jiang, Y.~Shi, and Z.~Ding, ``Federated learning via over-the-air
  computation,'' {\em IEEE Trans. Wireless Commun.}, vol.~19, no.~3,
  pp.~2022--2035, 2020.

\bibitem{alistarh2017qsgd}
D.~Alistarh, D.~Grubic, J.~Li, R.~Tomioka, and M.~Vojnovic, ``{QSGD}:
  Communication-efficient {SGD} via gradient quantization and encoding,'' in
  {\em Adv. Neural Info. Proc. Syst. (NIPS)}, (Long Beach, USA), Dec. 2017.

\bibitem{bernstein2018signsgd}
J.~Bernstein, Y.-X. Wang, K.~Azizzadenesheli, and A.~Anandkumar, ``sign{SGD}:
  Compressed optimisation for non-convex problems,'' in {\em Proc. Intl. Conf.
  Mach. Learning (ICML)}, (Stockholmsm{\"a}ssan, Stockholm Sweden), July 2018.

\bibitem{du2020high}
Y.~Du, S.~Yang, and K.~Huang, ``High-dimensional stochastic gradient
  quantization for communication-efficient edge learning,'' {\em IEEE Trans.
  Signal Process.}, vol.~68, pp.~2128--2142, 2020.

\bibitem{chang2020communication}
W.-T. Chang and R.~Tandon, ``Communication efficient federated learning over
  multiple access channels,'' {\em [Online]. Available:
  https://arxiv.org/pdf/2001.08737.pdf}, 2020.

\bibitem{zhu2020one}
G.~Zhu, Y.~Du, D.~Gunduz, and K.~Huang, ``One-bit over-the-air aggregation for
  communication-efficient federated edge learning: Design and convergence
  analysis,'' {\em [Online]. Available: https://arxiv.org/pdf/2001.05713.pdf},
  2020.

\bibitem{shokri2017membership}
R.~Shokri, M.~Stronati, C.~Song, and V.~Shmatikov, ``Membership inference
  attacks against machine learning models,'' in {\em IEEE Symp. Secur. and
  Privacy (SP)}, pp.~3--18, IEEE, 2017.

\bibitem{abadi2016deep}
M.~Abadi, A.~Chu, I.~Goodfellow, H.~B. McMahan, I.~Mironov, K.~Talwar, and
  L.~Zhang, ``Deep learning with differential privacy,'' in {\em Proc. ACM
  SIGSAC Conf. Comput. Commun. Secur.}, (Vienna, Austria), Oct. 2016.

\bibitem{wu2019value}
N.~Wu, F.~Farokhi, D.~Smith, and M.~A. Kaafar, ``The value of collaboration in
  convex machine learning with differential privacy,'' {\em [Online].
  Available: https://arxiv.org/pdf/1906.09679.pdf}, 2019.

\bibitem{agarwal2018cpsgd}
N.~Agarwal, A.~T. Suresh, F.~X.~X. Yu, S.~Kumar, and B.~McMahan, ``cp{SGD}:
  Communication-efficient and differentially-private distributed {SGD},'' in
  {\em Adv. Neural Info. Proc. Syst. (NIPS)}, (Montreal, Canada), Dec. 2018.

\bibitem{wei2020federated}
K.~Wei, J.~Li, M.~Ding, C.~Ma, H.~H. Yang, F.~Farokhi, S.~Jin, T.~Q. Quek, and
  H.~V. Poor, ``Federated learning with differential privacy: Algorithms and
  performance analysis,'' {\em IEEE Trans. Inf. Forensics Secur.}, 2020.

\bibitem{balle2018privacy}
B.~Balle, G.~Barthe, and M.~Gaboardi, ``Privacy amplification by subsampling:
  Tight analyses via couplings and divergences,'' in {\em Adv. Neural Info.
  Proc. Syst. (NIPS)}, (Montreal, Canada), Dec. 2018.

\bibitem{gandikota2019vqsgd}
V.~Gandikota, R.~K. Maity, and A.~Mazumdar, ``{vqSGD}: Vector quantized
  stochastic gradient descent,'' {\em [Online]. Available:
  https://arxiv.org/pdf/1911.07971.pdf}, 2019.

\bibitem{seif2020wireless}
M.~Seif, R.~Tandon, and M.~Li, ``Wireless federated learning with local
  differential privacy,'' {\em [Online]. Available:
  https://arxiv.org/pdf/2002.05151.pdf}, 2020.

\bibitem{koda2020differentially}
Y.~Koda, K.~Yamamoto, T.~Nishio, and M.~Morikura, ``Differentially private
  aircomp federated learning with power adaptation harnessing receiver noise,''
  {\em [Online]. Available: https://arxiv.org/pdf/2004.06337.pdf}, 2020.

\bibitem{sonee2020efficient}
A.~Sonee and S.~Rini, ``Efficient federated learning over multiple access
  channel with differential privacy constraints,'' {\em [Online]. Available:
  https://arxiv.org/pdf/2005.07776.pdf}, 2020.

\bibitem{zhao2018federated}
Y.~Zhao, M.~Li, L.~Lai, N.~Suda, D.~Civin, and V.~Chandra, ``Federated learning
  with non-iid data,'' {\em [Online]. Available:
  https://arxiv.org/pdf/1806.00582.pdf}, 2018.

\bibitem{wang2019adaptive}
S.~Wang, T.~Tuor, T.~Salonidis, K.~K. Leung, C.~Makaya, T.~He, and K.~Chan,
  ``Adaptive federated learning in resource constrained edge computing
  systems,'' {\em IEEE J. Sel. Areas Commun.}, vol.~37, no.~6, pp.~1205--1221,
  2019.

\bibitem{reisizadeh2020fedpaq}
A.~Reisizadeh, A.~Mokhtari, H.~Hassani, A.~Jadbabaie, and R.~Pedarsani,
  ``Fedpaq: A communication-efficient federated learning method with periodic
  averaging and quantization,'' in {\em Proc. Intl. Conf. Artif. Intell. Stat.
  (AISTATS)}, (Palermo, Italy), pp.~2021--2031, June 2020.

\bibitem{li2018federated}
T.~Li, A.~K. Sahu, M.~Zaheer, M.~Sanjabi, A.~Talwalkar, and V.~Smith,
  ``Federated optimization in heterogeneous networks,'' {\em [Online].
  Available: https://arxiv.org/pdf/1812.06127.pdf}, 2018.

\bibitem{ravi2019efficient}
S.~Ravi, ``Efficient on-device models using neural projections,'' in {\em Proc.
  Intl. Conf. Mach. Learning (ICML)}, (Long Beach, USA), pp.~5370--5379, June
  2019.

\bibitem{debaenst2020rms}
W.~Debaenst, A.~Feys, I.~Cui{\~n}as, M.~Garc{\'\i}a~S{\'a}nchez, and
  J.~Verhaevert, ``{RMS} delay spread vs. coherence bandwidth from {5G} indoor
  radio channel measurements at 3.5 {GHz} band,'' {\em Sensors}, vol.~20,
  no.~3, p.~750, 2020.

\bibitem{wang2018doppler}
S.~Wang, K.~Guan, D.~He, G.~Li, X.~Lin, B.~Ai, and Z.~Zhong, ``Doppler shift
  and coherence time of {5G} vehicular channels at 3.5 {GHz},'' in {\em Proc.
  IEEE Intl. Symp. on Antennas and Propagation \& USNC-URSI National Radio
  Science Meeting}, (Boston, USA), pp.~2005--2006, IEEE, July 2018.

\bibitem{cao2020optimized}
X.~Cao, G.~Zhu, J.~Xu, and K.~Huang, ``Optimized power control for over-the-air
  computation in fading channels,'' {\em IEEE Trans. Wireless Commun.}, 2020.

\bibitem{mahmood2019time}
A.~Mahmood, M.~I. Ashraf, M.~Gidlund, J.~Torsner, and J.~Sachs, ``Time
  synchronization in 5{G} wireless edge: Requirements and solutions for
  critical-mtc,'' {\em IEEE Commun. Mag.}, vol.~57, no.~12, pp.~45--51, 2019.

\bibitem{karimi2016linear}
H.~Karimi, J.~Nutini, and M.~Schmidt, ``Linear convergence of gradient and
  proximal-gradient methods under the polyak-lojasiewicz condition,'' in {\em
  Joint European Conf. on Mach. Learn. Knowl. Discovery in Databases (ECML
  KDD)}, (Riva del Garda, Italy), Sep. 2016.

\bibitem{bottou2018optimization}
L.~Bottou, F.~E. Curtis, and J.~Nocedal, ``Optimization methods for large-scale
  machine learning,'' {\em Siam Review}, vol.~60, no.~2, pp.~223--311, 2018.

\bibitem{friedlander2012hybrid}
M.~P. Friedlander and M.~Schmidt, ``Hybrid deterministic-stochastic methods for
  data fitting,'' {\em SIAM Journal on Scientific Computing}, vol.~34, no.~3,
  pp.~A1380--A1405, 2012.

\bibitem{zhao2015stochastic}
P.~Zhao and T.~Zhang, ``Stochastic optimization with importance sampling for
  regularized loss minimization,'' in {\em Proc. Intl. Conf. Mach. Learning
  (ICML)}, (Lille, France), July 2015.

\bibitem{chen2020understanding}
X.~Chen, Z.~S. Wu, and M.~Hong, ``Understanding gradient clipping in private
  sgd: A geometric perspective,'' {\em [Online]. Available:
  https://arxiv.org/pdf/2006.15429.pdf}, 2020.

\bibitem{zhang2019gradient}
J.~Zhang, T.~He, S.~Sra, and A.~Jadbabaie, ``Why gradient clipping accelerates
  training: A theoretical justification for adaptivity,'' in {\em Proc. Intl.
  Conf. Learning Representations (ICLR)}, (New Orleans, USA), May 2019.

\end{thebibliography}

\begin{IEEEbiography}
[{\includegraphics[width=1in,clip,keepaspectratio]{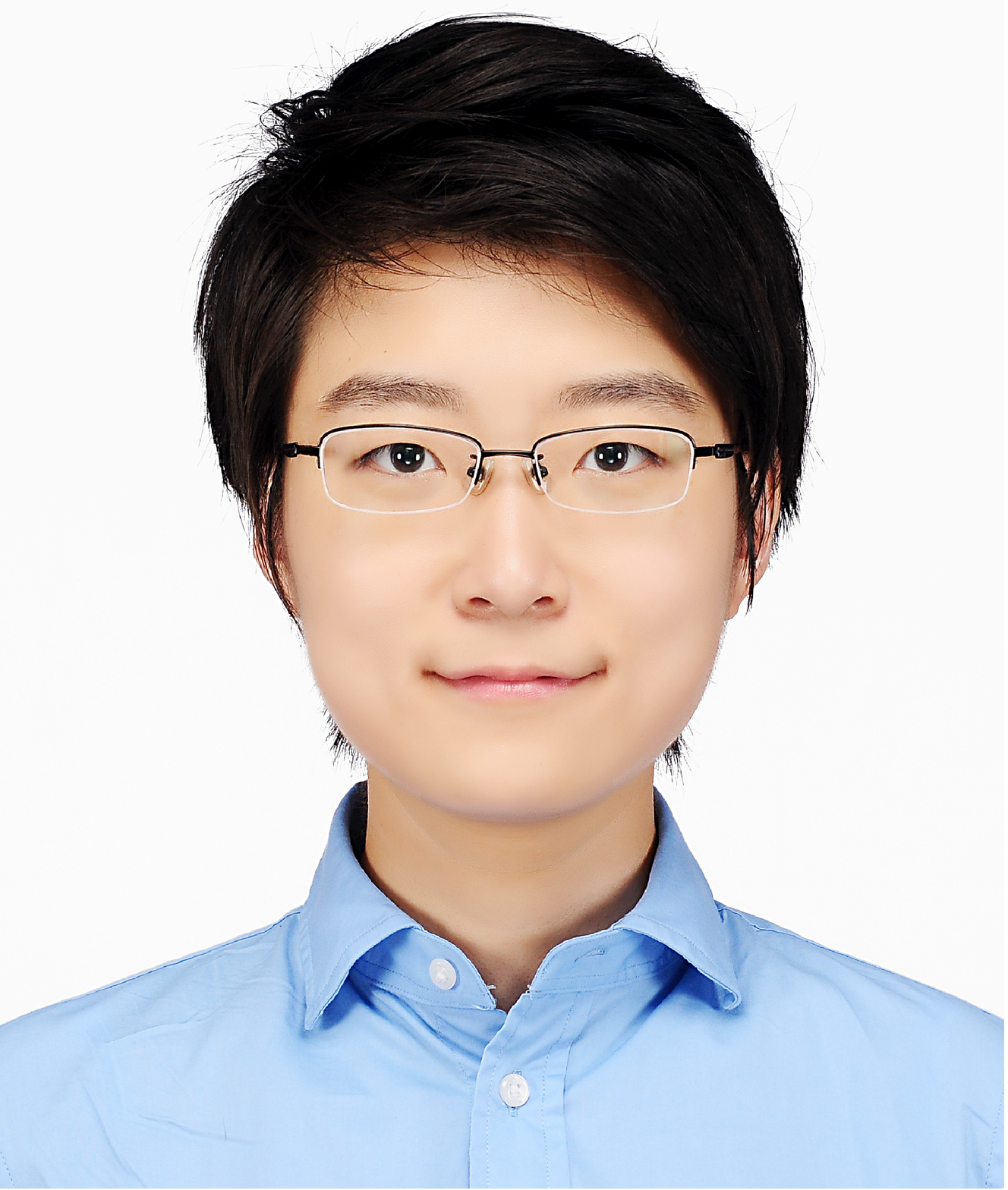}}]
{Dongzhu Liu} received the B.Eng. degree from the University of Electronic Science and Technology of China (UESTC) in 2015, and the Ph.D. degree from The University of Hong Kong in 2019. She is now a postdoctoral research associate in the Dept. of  Engineering at King's College London.  Her research interests include edge intelligence, federated learning, and wireless communications.
\end{IEEEbiography}

\begin{IEEEbiography}
[{\includegraphics[width=1in,clip,keepaspectratio]{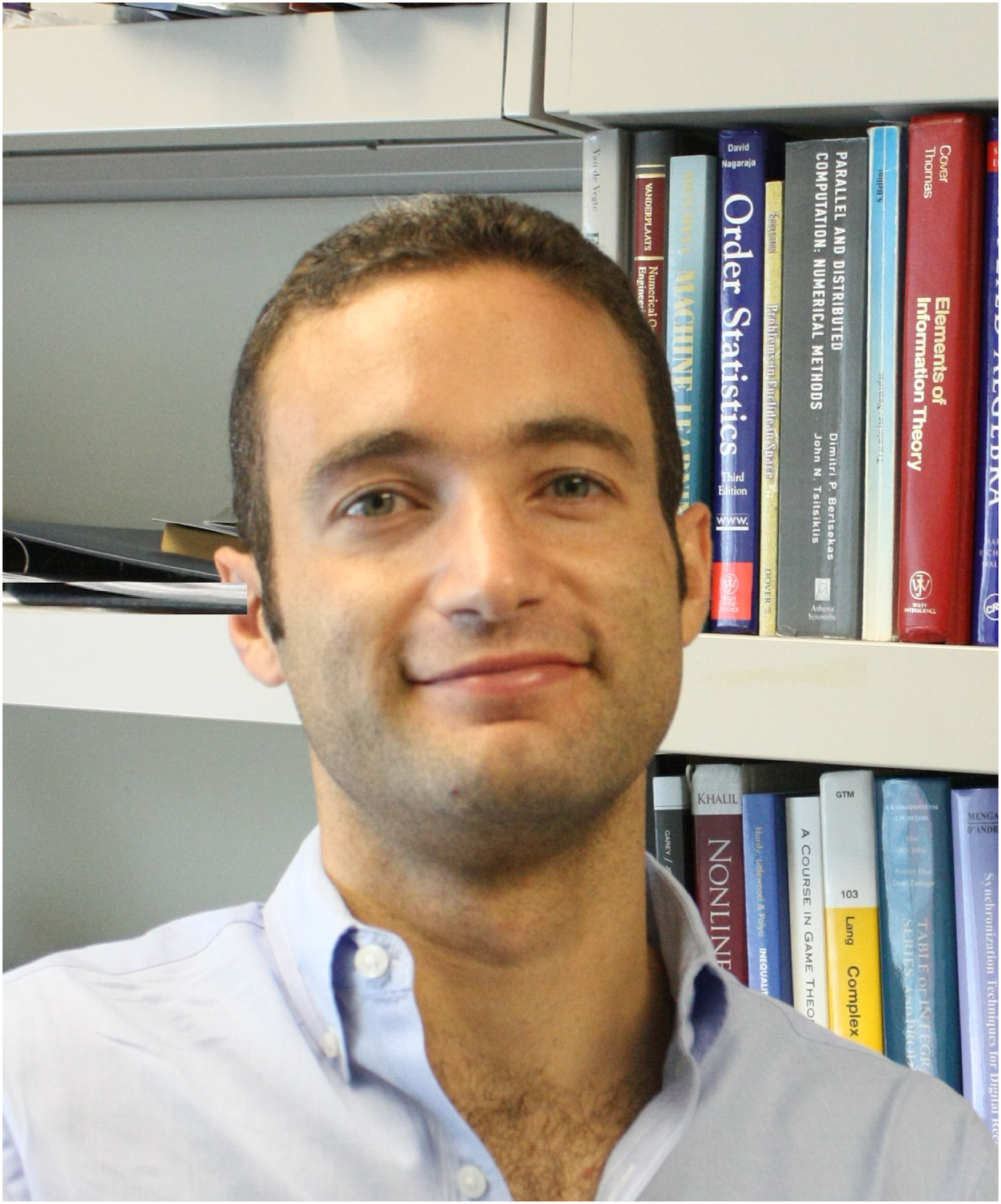}}]
{Osvaldo Simeone} (Fellow, IEEE) received an M.Sc. degree (with honors) and a Ph.D. degree in information engineering from Politecnico di Milano, Milan, Italy, in 2001 and 2005, respectively. Presently, he is a Professor of Information Engineering with the Centre for Telecommunications Research at the Department of Engineering of King's College London, where he directs the King's Communications, Learning and Information Processing lab.  From 2006 to 2017, he was a faculty member of the Electrical and Computer Engineering (ECE) Department at New Jersey Institute of Technology (NJIT), where he was affiliated with the Center for Wireless Information Processing (CWiP). His research interests include information theory, machine learning, wireless communications, and neuromorphic computing. 

Dr Simeone is a co-recipient of the 2019 IEEE Communication Society Best Tutorial Paper Award, the 2018 IEEE Signal Processing Best Paper Award, the 2017 JCN Best Paper Award, the 2015 IEEE Communication Society Best Tutorial Paper Award and of the Best Paper Awards of IEEE SPAWC 2007 and IEEE WRECOM 2007. He was awarded a Consolidator grant by the European Research Council (ERC) in 2016. His research has been supported by the U.S. NSF, the ERC, the Vienna Science and Technology Fund, as well as by a number of industrial collaborations. He currently serves in the editorial board of the IEEE Signal Processing Magazine and is the vice-chair of the Signal Processing for Communications and Networking Technical Committee of the IEEE Signal Processing Society. He was a Distinguished Lecturer of the IEEE Information Theory Society in 2017 and 2018. Dr Simeone is a co-author of two monographs, two edited books published by Cambridge University Press, and more than one hundred research journal papers. He is a Fellow of the IET and of the IEEE.

\end{IEEEbiography}

\end{document}